\documentclass[structabstract]{aa}  
\usepackage{graphicx}
\begin{document}
\title{A critical analysis of the UV Luminosity Function at redshift $\sim$7
from deep WFC3 data}

   \author{A. Grazian \inst{1}
          \and
          M. Castellano \inst{1}
          \and
          A. M. Koekemoer \inst{2}
          \and
          A. Fontana \inst{1}
          \and
          L. Pentericci \inst{1}
          \and
	  V. Testa \inst{1}
	  \and
          K. Boutsia \inst{1}
	  \and
          E. Giallongo \inst{1}
	  \and
          M. Giavalisco \inst{3}
	  \and
          P. Santini \inst{1}
          }

   \offprints{A. Grazian, \email{grazian@oa-roma.inaf.it}}

\institute{INAF - Osservatorio Astronomico di Roma, Via Frascati 33,
I--00040, Monteporzio, Italy
\and Space Telescope Science Institute, 3700 San Martin Drive, Baltimore,
MD 21218
\and Department of Astronomy, University of Massachusetts, 710 North
Pleasant Street, Amherst, MA 01003
}

   \date{Received 14/09/2010; accepted 18/04/2011}

   \authorrunning{Grazian et al.}
   \titlerunning{A critical analysis of the $z\sim 7$ UV LF}

% \abstract{}{}{}{}{} 
% 5 {} token are mandatory
 
  \abstract
   {
The study of the Luminosity Function (LF) of Lyman Break Galaxies (LBGs)
at z=7 is very important for ascertaining their role in the
reionization of the Universe. These galaxies can be used also to investigate
in detail the processes of formation and evolution of galactic structures
in the infancy of our Universe.
   }
   {
In this work we plan to perform
a detailed and critical analysis of the statistical and systematic errors
in the $z\sim 7$ LF determination.
   }
   {
To this aim, we have assembled a large sample of candidate LBGs at
$z\sim 7$ from different surveys, spanning a large variety of areas and
depths. In particular, we have combined data from the deep ($J<27.4$)
and ultradeep ($J<29.2$) surveys recently acquired with the new WFC3 NIR
camera on HST, over the GOODS-ERS ($\sim$40 sq. arcmin.) and the HUDF
($\sim$4 sq. arcmin.) fields,
with ground based surveys in wide and shallow areas from
Hawk-I@VLT and HyperSuprimecam@Subaru.
We have used public ACS
images in the z band to select z-dropout galaxies, and other public
data both in the blue (BVI) and in the red bands to reject
possible low-redshift interlopers.
We have compared our results with
extensive Monte Carlo simulations to quantify the observational
effects of our selection criteria as well as the effects of
photometric scatter, color selections or the morphology of the
candidates.
   }
   {
We have found that the number density of faint LBGs at $z\sim 7$ is
only marginally sensitive to the color selection adopted, but it is strongly
dependent from the assumption made on the half light distributions
of the simulated galaxies, used to correct the observed sample for
incompleteness. The slope of the faint end of the LBGs LF has thus a rather
large uncertainty, due to the unknown distribution of physical sizes of
the $z\sim 7$ LBGs. The implications of these uncertainties have
been neglected by previous works.
   }
{
We conclude that galaxies at $z\sim 7$ are unable to reionize the
Universe unless there is a significant evolution in the clumpiness of
the IGM or in the escape fraction of ionising photons or, alternatively,
there is a large population of $z\sim 7$ LBGs with large physical dimensions
but still not detected by the present observations.
}


   \maketitle
%
%________________________________________________________________

\section{Introduction}

The detailed understanding of
the physical properties of the first building blocks of present-day
galaxies is fundamental to set up the zero point for the theoretical
models dealing with the formation and evolution of galactic structures
in the Universe. In particular, the study of the faint end of the LBG
LF at z=7 is very important for
understanding their contribution to the reionization processes of the
high-z Universe.

At $z\sim 6$ the Universe is almost re-ionized (\cite{fan06,becker07})
and there is both observational (\cite{meiksin09,songaila10,komatsu10})
and theoretical (\cite{gnedin06,cen10})
mounting evidence that the crucial transformation from a neutral to an ionized
state should happen around $z\sim 9$. The exact timescale of the
reionization process, however, is not clear, the alternatives going from
an extended period ($z\sim 7-11$, \cite{dunkley09}) to a more sudden
transition (\cite{cen03}).

The new Subaru (\cite{ouchi}), Hawk-I (\cite{castellano09,castellano10},
hereafter C10a and C10b),
and WFC3 (\cite{bouwens10c}) surveys have started investigating the
properties of galaxies during the reionization epoch, enlarging the
small sample of galaxy candidates at $z\sim 7$ sketched by
NICMOS (\cite{bouwens06,Mannucci2007,hudf09}). Recently,
\cite{oesch09} derived the number density of faint LBGs at z=7 in the
HUDF, concluding that these low luminosity galaxies are enough to
reionize the Universe, assuming that the IMF, the clumpiness of the IGM,
the escape fraction of ionising photons and their metallicity have not
significantly evolved with
respect to their local properties. Using the same data-set,
\cite{bunker09} and \cite{mclure10} found that it would be difficult for
the observed
population of high-redshift star-forming galaxies to achieve
reionization by $z\sim 6-7$ without a significant contribution from
galaxies well below the detection limits, together with significant
variations in the escape fraction of ionising photons. Quantifying
the number density of $z\ge 7$ galaxies is therefore critical in order
to check whether there are sources responsible for the re-ionization,
such as Pop III dominated primordial galaxies, mini-black holes or others
(\cite{Venkatesan2003,Madau2004}).

The different results obtained for the LF of LBGs at
z=7 (C10a, C10b, \cite{ouchi,capak,oesch09,mclure10,bunker09,wilkins,
hickey,yan,finkelstein})
should be ascribed both to the different data-sets used and to the
different techniques adopted to analyse the data. We want to study in
detail the possible systematics/uncertainties in the LF estimate, in
particular due to the choice of the color cut adopted, the morphology of the
z=7 galaxies assumed during simulations, and the statistical tools used
in the LF derivation. Moreover, extending the LF at z=7 down to faint
magnitude limits is fundamental to break current degeneracies between
$M^*$ and $\Phi^*$, and to put strong constraints on the number density of
faint LBGs at high-z. Here we combine a re-analysis of the extremely deep
WFC3 observations (\cite{oesch09}) with the selection of bona-fide $z\sim
7$ LBGs in wide areas of the sky (C10a, C10b, \cite{ouchi}) in order to derive
stringent constraint to the reionization process at $z\sim 7$.

The paper is organised as follows. Section 2 and 3 describe the imaging data
used, the photometric analysis, and the
selection of the $z\sim 7$ LBG candidates. The simulations
carried out to derive incompleteness and contamination from lower-z galaxies
are discussed in Section 4.
In Section 5 we discuss the systematics affecting the LF (color criteria and
morphology of the galaxies adopted during the simulations),
with our results on the LF derivation.
Section 6 and 7 are devoted to our discussions and conclusions.

Throughout the whole paper, observed and rest--frame magnitudes are in
the AB system, and we adopt the $\Lambda$-CDM concordance model
($H_0=70km/s/Mpc$, $\Omega_M=0.3$ and $\Omega_{\Lambda}=0.7$).

%__________________________________________________________________

\section{Data and Photometry}

We have analysed two different data-sets observed with the new NIR camera of
HST, the Wide Field Camera 3 (WFC3\footnote{http://www.stsci.edu/hst/wfc3}):
the Early Release Science on the GOODS-S field (GOODS-ERS) and the Hubble
Ultra Deep Field (HUDF) programs.

\subsection{ERS}

The GOODS-ERS WFC3/IR dataset is a total of 60 HST orbits consisting of
10 contiguous pointings in the GOODS-South field (HST Program ID 11359),
using 3 filters per visit ($Y_{098}$, $J_{125}$, $H_{160}$), and 2 orbits
per filter (for a total of 4800-5400s per pointing and filter).
The total area covered by the GOODS-ERS is $\sim 40 sq. arcmin.$
down to Y=27.3, J=27.4, and H=27.4 magnitudes at $5\sigma$ in an area of
$\sim 0.11$ $arcsec^2$ (corresponding to 2 times the FWHM of the images).
The ERS data were acquired and used to demonstrate the WFC3 capabilities
on the study of low-z and high-z galaxies (\cite{windhorst}), and
the raw images were publicly released by the instrument
team. Thanks to its fine pixel sampling and high sensitivity in the NIR,
WFC3 is an ideal instrument to study the $z\ge 7$ galaxy populations.
The raw images of the ERS dataset have been reduced and used to search for
zdrop candidates by \cite{wilkins,wilkins2} and \cite{bouwens10c}.

The WFC3 ERS images were reduced using the Multidrizzle software
(\cite{multidrizzle}), adopting on-orbit SPARS100 darks and also
correcting for the gain differences between different quadrants.
Large-scale residuals on each exposure which are apparently due
to scattered light have been removed, and the satellite trails have
been masked on affected exposures. The astrometry of the images is
bound to the GOODS-S v2.0 z-band (\cite{giavalisco04}). The matching is
robust even for a single exposure, with 400-600 sources matching, so
that the overall alignment seems to be robust to a level of $\sim
10mas$. The pixel size of the reduced images is 0.06 arcsec
(a factor of 2 rebinning was applied with respect to the
original pixel size of 0.135$\times$0.121 arcsec$^2$), with a PSF
of 0.18 arcsec.

\subsection{HUDF}

The HUDF WFC3/IR dataset (HST Program ID 11563) is a total of 60 HST
orbits in a single pointing (\cite{oesch09,bouwens10})
in three broad-band filters
(16 orbits in $Y_{105}$, 16 in $J_{125}$, and 28 in $H_{160}$). It is
the deepest NIR image ever taken, reaching Y=29.3, J=29.2, and
H=29.2 total magnitudes for point like sources at 5 sigma
(this S/N ratio is computed
in an aperture of $\sim 0.11$ $arcsec^2$, corresponding to one FWHM).
The area covered by the WFC3-HUDF imaging is 4.7 $sq. arcmin.$,
and the IR data have been drizzled to the ACS-HUDF data (\cite{udf}),
with a resulting pixel scale of 0.03 arcsec, with a FWHM
of 0.18 arcsec.

The HUDF images were reduced in the same way as the ERS dataset, i.e.
background effects are properly corrected (on-orbit darks, quadrant
gain corrections, multiplicative sky-flats and additive low-level
residuals on large scales all removed), and the images have been
astrometrically matched using the ACS HUDF sources.
For the $Y_{105}$ filter, we have selected a subset of 7 visits out of 9
that are not badly affected by persistence from bright targets
observed before these observations; the images in the other two
filters (J and H) are not affected by this persistence problem.
The magnitude limit of the cleaned Y-band image is 29.0, at 5 sigma
in $\sim 0.11$ $arcsec^2$. The GOODS-ERS and HUDF datasets will be described
in detail in Koekemoer et al. (2011, in prep.).

\subsection{Photometry}

The photometric catalog of the ERS and HUDF fields have been derived
in a consistent way. Galaxies have been detected in the $J_{125}$ band,
and their total magnitudes have been computed using the MAG\_BEST of
SExtractor (\cite{sex}), for galaxies more extended than 0.11 $arcsec^2$, while
circular aperture photometry (diameter equal to 2 times the WFC3
FWHM) is used for smaller sources. The magnitudes which are computed
in a circular aperture have been corrected to total magnitude by
applying a correction of 0.4 magnitudes. Colors in BVIZYH have
been measured running SExtractor in dual image mode, using isophotal
magnitudes for all the galaxies, independently from their area. Since
the FWHM of the ACS images ($0.12 \arcsec$) is better than the WFC3
ones, we smoothed the ACS bands with an appropriate kernel to reproduce
the resolution of the NIR WFC3 images. This ensures both precise
colors for extended objects and non-biased photometry for faint
sources.

As a comparison, we produced, only for the HUDF field, an equivalent
BVIZYJH multiwavelength catalog, using the SExtractor aperture magnitudes
instead of the isophotal magnitudes to derive galaxy colors. In the resulting
Z-Y vs Y-J plot, indeed, galaxies are much more scattered than in
Fig.\ref{fig:hudfcand}, and the bulk of low-z galaxies are scattered to
$Z-Y\sim 1$ when the aperture colors have been used, while it is limited to
$Z=Y\sim 0.5$ if isophotal colors are adopted. This is due to the fact that
the resulting isophotal areas are smaller than the circular apertures
adopted, and thus are less noisy.

%__________________________________________________________________

\section{Selection of z=7 candidates: color criteria}

\begin{figure}
\includegraphics[width=9cm]{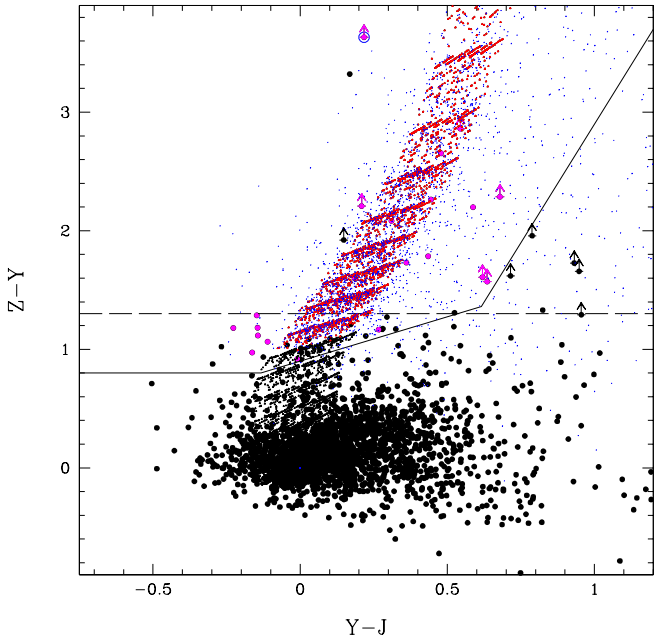}
\caption{
The $Z-Y$ vs $Y-J$ color-color diagram for the HUDF dataset limited at
$J<29.2$. Big points are observed galaxies, small black points are CB07 models
at $0.0<E(B-V)<0.3$, with small red points those at $z>6.5$. The small blue
points are the result of our simulations, with synthetic objects inserted in
the FITS images and recovered using the same procedure used for the observed
catalog. The solid line is the color cut of the A criterion, while the
dashed line shows the B criterion (described in Sect. 3). Candidate galaxies at
$z\sim 7$ selected with criterion C are shown in magenta.
Vertical arrows are upper limits in
the $Z$ band, while the blue circle at $Z-Y>3.6$ is the suspected SNa
outlined by \cite{oesch09}.
}
\label{fig:hudfcand}
\end{figure}

The selection of galaxies at $z\sim 7$ uses the well known ``drop-out''
or ``Lyman-break'' technique. At $6.5<z<7.5$, this feature is sampled
by the large $Z-Y$ color, as shown in Figure~\ref{fig:hudfcand} for the
HUDF field.

Recently, a number of slightly different color criteria have been used
to select
bona fide galaxies at $z\sim 7$, rejecting lower-z objects and cold stars
entering the z-dropout criteria due to photometric scatter. Despite the
differences in these criteria, however, the results obtained i.e. by
C10a,C10b,\cite{wilkins,bunker09,oesch09} are comparable, at least at the
bright luminosities.
We aim here at applying the different color criteria to check whether the
LF, which is derived from the data after detailed simulations, is robust
against all the random and systematic effects that can affect each
selection criterion.

With WFC3 data (ERS and HUDF), we explore three different color
criteria, one of \cite{oesch09}, one of \cite{bunker09}, and one
proposed by us:
\begin{itemize}
\item A)
The \cite{oesch09} criterion for the HUDF is:
\begin{eqnarray*}
z-Y_{105}&>&0.8,\\
z-Y_{105}&>&0.9+0.75(Y_{105}-J_{125}),\\
z-Y_{105}&>&-1.1+4.0(Y_{105}-J_{125}),\\
S/N(J_{125})&>&5, ~ S/N(Y_{105})>5,\\
S/N(V)&<&2, ~ S/N(I)<2.
\end{eqnarray*}

For the ERS, not analysed in \cite{oesch09}, an equivalent criterion has
been adopted here, that takes into account the difference between the
$Y_{098}$ filter instead of the $Y_{105}$ one:
\begin{eqnarray*}
z-Y_{098}&>&1.1,\\
z-Y_{098}&>&0.55+1.25(Y_{098}-J_{125}),\\
z-Y_{098}&>&-0.5+2.0(Y_{098}-J_{125}),\\
S/N(J_{125})&>&5, ~ S/N(Y_{098})>5,\\
S/N(V)&<&2, ~ S/N(I)<2.
\end{eqnarray*}

For the HUDF field, the $S/N(J)>5$ criterion corresponds to $J<29.0$,
while for the ERS field it allows the selection of galaxies brighter
than $J=27.4$.\\

\item B)
The selection criterion of \cite{bunker09} is:
\begin{eqnarray*}
z-Y&>&1.3,\\
S/N(V)&<&2,\\
S/N(I)&<&2.
\end{eqnarray*}

The samples of \cite{bunker09} and \cite{wilkins,wilkins2} are
limited to $Y=28.5$ for the HUDF and $Y=27.2$ for the ERS field,
respectively, aiming
at a more robust selection of $z\sim 7$ candidates. We adopt the same
magnitude limits in this case.\\

\item C)
We propose a modification of the \cite{oesch09} criterion, based on
simple considerations. We adopt the same color criteria:
\begin{eqnarray*}
z-Y_{105}&>&0.8,\\
z-Y_{105}&>&0.9+0.75(Y_{105}-J_{125}),\\
z-Y_{105}&>&-1.1+4.0(Y_{105}-J_{125})
\end{eqnarray*}
for the HUDF and
\begin{eqnarray*}
z-Y_{098}&>&1.1,\\
z-Y_{098}&>&0.55+1.25(Y_{098}-J_{125}),\\
z-Y_{098}&>&-0.5+2.0(Y_{098}-J_{125})
\end{eqnarray*}
for the ERS.

To avoid contamination of faint spurious sources, we add the criterion
$J_{125}-H_{160}\ge -1.0$.
This additional criterion removes only two extremely faint galaxies
with $29.0<J_{125}<29.2$ magnitudes, which are detected only in J
but not in the H band. Indeed, the observed $J-H$ color for our z-dropout
candidates shows a bimodal distribution, with a main peak at
$J_{125}-H_{160}\sim -0.2$ and two isolated sources at
$J_{125}-H_{160}\sim -1.2$. Since LBGs are expected to be almost flat
at $\lambda\sim 2000$\AA~ rest frame, we interpret these source
with $J_{125}-H_{160}\le -1.0$ as spurious
contaminants and remove them from the selected sample.
As we have verified in the following
(see section 5.1), this further selection
does not change the resulting luminosity function at $z\sim 7$.

For the non-detection in bands bluer than
$Z$, we adopt the same criteria used in \cite{castellano09,castellano10}
($S/N<2$ in all BVI
bands and $S/N<1$ in at least two of them). We apply however only a
single cut in the J magnitude ($J<29.2$ for the HUDF and $J<27.4$ for
the ERS), since this band corresponds to 1500\AA~ rest frame.
We avoid the double
selection $S/N(J)>5$ and $S/N(Y)>5$, as adopted by \cite{oesch09},
since it could be less sensitive to red galaxies in the $Y-J$ color
at faint $J$ magnitudes.
\end{itemize}

Fig.\ref{fig:hudfcand} shows the Z-Y vs Y-J color plot for the HUDF field.
The solid segments show the A and C criteria, while the criterion B is
indicated by the dashed line.

We select 13, 6, and 20 candidates following the A, B, and C criteria
in the HUDF field, while in the ERS we find 14, 6, and 22 z-dropout
candidates, respectively. The small numbers for the candidates
provided by the B criterion reflects the more conservative
cut adopted by \cite{bunker09} to get rid of the lower-z contaminants
with $z-Y\sim 1$. We have excluded from these candidates the bright
transient (probably a SNa), outlined by \cite{oesch09} in the HUDF
field (blue circle in Fig.\ref{fig:hudfcand} at $Z-Y>3.6$).
This relatively bright transient
was also identified by \cite{mclure10} and \cite{bunker09},
while \cite{yan11} showed its possible light curve.

Table \ref{table:hudf} and \ref{table:ers} summarize the properties
of the z-dropout candidates in the HUDF and ERS field, respectively.

%_____________________________________________________________
%                                             Two column Table
\begin{table*}
\caption{Galaxy candidates at $z\sim 7$ in the HUDF field}
\label{table:hudf}
\centering
\begin{tabular}{l c c c c c c c c c}     % 10 columns
\hline\hline
id &  RAD &         DEC &          J &    z-Y & Y-J & J-H & rh & Sel & References\\
\hline
HUDF-1095 & 03:32:42.561 & -27:46:56.60 & 26.07 & 1.60 &  0.19 & -0.02 & 0.256 & A,B,C & 1,2,3,4,6,7,8,9,10,11,12,13,14\\
HUDF-1344 & 03:32:38.805 & -27:47:07.16 & 27.09 & 2.66 &  0.47 &  0.07 & 0.147 & A,B,C & 1,2,3,4,5,6,8,9,10,13\\
HUDF-566  & 03:32:42.562 & -27:47:31.44 & 27.18 & 1.73 &  0.36 & -0.22 & 0.138 & A,B,C & 1,3,4,6,8,9,10,11,13,14\\
HUDF-1537 & 03:32:41.048 & -27:47:15.59 & 27.36 & 2.08 &  0.31 & -0.35 & 0.134 & A,B,C & 6,8,10,11\\
HUDF-819  & 03:32:44.702 & -27:46:44.29 & 27.47 & 2.86 &  0.55 & -0.13 & 0.161 & A,B,C & 6,8,9,10,11,14 (a)\\
HUDF-1574 & 03:32:39.554 & -27:47:17.49 & 27.62 & 2.19 &  0.59 &  0.04 & 0.120 & A,B,C & 1,3,6,8,9,10,11\\
HUDF-684  & 03:32:43.135 & -27:46:28.47 & 27.82 & 2.26 &  0.44 &  0.10 & 0.168 & A,C & 6,8,9,10,11,13,14\\
HUDF-101  & 03:32:37.214 & -27:48:06.15 & 27.90 & 2.29 &  0.68 & -0.20 & 0.144 & A,C & 6,8,9,10,11\\
HUDF-1554 & 03:32:36.378 & -27:47:16.24 & 27.94 & 1.07 & -0.11 & -0.22 & 0.160 & A,C & 6,8,9,10,11,13,14\\
HUDF-1828 & 03:32:43.782 & -27:46:33.69 & 27.99 & 0.98 & -0.17 & -0.10 & 0.160 & A,C & 8,11\\
HUDF-1097 & 03:32:39.578 & -27:46:56.47 & 28.07 & 1.29 & -0.15 & -0.45 & 0.154 & A,C & 6,8,10,11,14\\
HUDF-2373 & 03:32:38.360 & -27:46:11.90 & 28.07 & 1.79 &  0.43 & -0.27 & 0.140 & C & 9,10,11\\
HUDF-976  & 03:32:37.444 & -27:46:51.29 & 28.12 & 1.11 & -0.14 & -0.10 & 0.133 & A,C & 6,8,10,11,13,14\\
HUDF-2020 & 03:32:39.725 & -27:46:21.34 & 28.19 & 2.20 &  0.21 & -0.15 & 0.190 & A,C & 6,8,9,10,11,14\\
HUDF-421  & 03:32:37.796 & -27:47:40.44 & 28.21 & 0.92 & -0.01 &  0.14 & 0.169 & C & 6,8\\
HUDF-2347 & 03:32:41.824 & -27:46:11.26 & 28.34 & 1.16 &  0.27 & -0.17 & 0.180 & C & 10,11\\
HUDF-805  & 03:32:40.567 & -27:46:43.55 & 28.34 & 1.18 & -0.14 & -0.22 & 0.148 & C & 6,8,9,10,11,13,14\\
HUDF-1276 & 03:32:41.596 & -27:47:04.47 & 28.49 & 1.18 & -0.23 & -0.15 & 0.152 & C & 14\\
HUDF-1102 & 03:32:33.130 & -27:46:54.47 & 28.73 & 1.57 &  0.64 & -0.16 & 0.129 & C & -\\
HUDF-133  & 03:32:40.334 & -27:48:02.61 & 29.03 & 1.61 &  0.62 & -0.15 & 0.109 & C & -\\
\hline
\end{tabular}
\\
References: [1] Bouwens et al. (2004), [2] Bouwens \& Illingworth (2006),
[3] Labb\'e et al. (2006), [4] Bouwens et al. (2008), [5] Oesch et al. (2009),
[6] Oesch et al. (2010), [7] Fontana et al. (2010),
[8] McLure et al. (2010),
[9] Bunker et al. (2010), [10] Yan et al. (2010),
[11] Finkelstein et al. (2010), [12] Castellano et al. (2010a),
[13] Wilkins et al. (2011), [14] Bouwens et al. (2011).\\
Note (a): both [8] and [14] indicate a nearby object 03:32:44.74 -27:46:44.92.
\end{table*}
%

%_____________________________________________________________
%                                             Two column Table
\begin{table*}
\caption{Galaxy candidates at $z\sim 7$ in the ERS field}
\label{table:ers}
\centering
\begin{tabular}{l c c c c c c c c c}     % 10 columns
\hline\hline
id &  RAD &         DEC &          J &    z-Y & Y-J & J-H & rh & Sel & References\\
\hline
ERS-5659  & 03:32:22.658 & -27:43:00.64 & 25.54 & 1.34 &  0.25 & -0.01 & 0.202  & A,B,C &  1,2,3,5\\
ERS-3881  & 03:32:25.284 & -27:43:24.20 & 25.98 & 2.07 &  0.11 &  0.14 & 0.254  & A,B,C &  4,5\\
ERS-7034  & 03:32:24.095 & -27:42:13.91 & 26.08 & 1.22 &  0.38 & -0.17 & 0.275  & C &  4,5,6\\
ERS-285   & 03:32:10.407 & -27:45:40.80 & 26.43 & 1.13 & -0.01 &  0.17 & 0.225  & C &  -\\
ERS-1819  & 03:32:06.826 & -27:44:22.18 & 26.61 & 1.26 & -0.07 & -0.30 & 0.135  & A,C &  6\\
ERS-2200  & 03:32:22.934 & -27:44:09.92 & 26.72 & 1.75 &  0.10 &  0.01 & 0.128  & A,B,C &  5\\
ERS-3719  & 03:32:15.400 & -27:43:28.61 & 26.98 & 1.31 & -0.08 & -0.33 & 0.120  & A,B,C &  -\\
ERS-4682  & 03:32:23.155 & -27:42:04.69 & 27.00 & 1.88 &  0.09 & -0.61 & 0.178  & A,C &  5\\
ERS-10069 & 03:32:22.518 & -27:41:17.34 & 27.01 & 1.22 & -0.01 & -0.10 & 0.129  & A,C &  6\\
ERS-1676  & 03:32:08.359 & -27:44:27.40 & 27.03 & 1.95 &  0.29 &  0.01 & 0.159  & A,C &  -\\
ERS-457   & 03:32:10.035 & -27:45:24.53 & 27.09 & 3.70 &  0.02 & -0.47 & 0.146  & A,C &  -\\
ERS-8656  & 03:32:36.313 & -27:40:14.99 & 27.11 & 2.34 &  0.12 &  0.07 & 0.197  & A,C &  -\\
ERS-3875  & 03:32:09.853 & -27:43:24.01 & 27.13 & 3.97 & -0.29 &  0.21 & 0.185  & A,B,C &  -\\
ERS-6532  & 03:32:31.227 & -27:42:25.17 & 27.14 & 1.86 & -0.25 & -0.88 & 0.136  & A,B,C &  -\\
ERS-3642  & 03:32:41.793 & -27:43:30.05 & 27.16 & 1.59 &  0.46 & -0.12 & 0.357  & A,C &  -\\
ERS-4752  & 03:32:39.829 & -27:42:40.90 & 27.19 & 1.69 &  0.67 & -0.38 & 0.288  & C &  -\\
ERS-7187  & 03:32:33.860 & -27:42:09.73 & 27.30 & 3.19 &  0.34 & -0.11 & 0.140  & C &  -\\
ERS-148   & 03:32:07.287 & -27:45:54.29 & 27.31 & 2.90 & -0.18 & -0.63 & 0.130  & A,C &  -\\
ERS-4639  & 03:32:13.405 & -27:42:30.89 & 27.33 & 2.50 &  0.77 & -0.02 & 0.168  & C &  -\\
ERS-5945  & 03:32:16.767 & -27:43:07.16 & 27.35 & 3.02 &  0.25 & -0.22 & 0.145  & C &  -\\
ERS-5383  & 03:32:29.458 & -27:42:54.27 & 27.36 & 2.09 &  0.54 & -0.66 & 0.136  & C &  -\\
ERS-1965  & 03:32:14.988 & -27:44:17.54 & 27.37 & 2.82 &  0.04 & -0.67 & 0.148  & C &  -\\
\hline
\end{tabular}
\\
References: [1] Bouwens et al. (2010b), [2] Hickey et al. (2010),
[3] Castellano et al. (2010a), [4] Wilkins et al. (2010),
[5] Wilkins et al. (2011), [6] Bouwens et al. (2011).
\end{table*}

The number counts of the selected $z>7$ candidates are different for the three
different criteria. In the next paragraphs we will check if the
simulations, by including the different selection effects,
are able to correct for these differences, and if the resulting
Luminosity Function depends on the color criteria adopted.

\section{Simulations}

While the selection criteria described above are formally designed to select
a pure sample of high-z candidates, they are in practice
applied to very faint objects, typically close to the limiting depth
of the images. At these limits, systematics may significantly affect
their detection and the accurate estimate of their large color terms.
To take into account all the systematic effects (completeness,
photometric scatter) involved in the LF estimate, we carried out a set
of detailed simulations, with the aim of deriving the LF.

\subsection{The Stepwise method to compute the LF}

This method is similar to that used in
\cite{bouwens08}, improved to take into account the photometric scatter,
as described in detail in C10a and C10b. We
first produce a set of $(M_{1500},z)$ according to a flat distribution
both in redshift and in absolute magnitudes, spanning the redshift
range $5.5<z<8.5$ and the absolute magnitude interval $-25\le M_{1500}\le
-16$. We then convert the absolute magnitudes into observed magnitudes
convolving the synthetic SEDs of the LBGs through the relevant filter response
curves. For this purpose we have used the models of
\cite{Bruzual2007} (hereafter CB07) with the
following range of free parameters: metallicity: 0.02, 0.2, and 1
$Z_\odot$; age from 0.01 Gyr to the maximal age of the Universe at a
given $z$; E(B-V) from 0 to 0.3 following a \cite{Calzetti2000}
extinction law. Since Lyman-$\alpha$ emission has an important
influence on the selection function of Lyman break galaxies, as
shown in lower redshift samples
(e.g. \cite{Stanway2008,Dow2007}), we explicitly take into account
its effect by considering a Gaussian distribution of Ly-$\alpha$ rest-frame
equivalent width with FWHM of 30 \AA~ and centered at EW=0\AA.
We have also added the
intergalactic absorption using the average evolution as in
\cite{Madau1995}.

These synthetic galaxies are finally placed at random positions on the
real images.
We then recover the measured magnitudes in the synthetic images using
SExtractor with the same parameters adopted for the real images.
We simulate all the available bands, i.e. from the B to the H bands
for the ERS and HUDF
fields. To avoid an excessive and unphysical crowding in the
simulated images, we have included only 200 objects of the same flux
and morphology each time, after masking the regions of the images
where real objects have been detected. We repeated the simulation
until a total of at least $2\times10^4$ objects were tested for
each WFC3 field and for each morphological template adopted.
These simulations provide the transfer function between the input absolute
magnitudes and the observed colors.
In addition, the simulations can be used
to evaluate the uncertainties in the estimate of the color criteria
that we shall exploit to detect $z>6.5$ candidates, especially in the
$Z-Y$ and $Y-J$ combination, which is essential for a clear identification
of $z\sim 7$ galaxies.
These simulations indeed can be used to estimate the impact of different
systematics in the study of the LF. For example, since the Y band filter
used in the ERS field ($Y_{098}$) has a bluer effective wavelength than
the one in the HUDF field ($Y_{105}$), it turns out that the selection function
in redshift in these two fields is different. These systematics have been taken
fully into account with these simulations.

The simulations described above have been used to derive an estimate
of the $z\sim 7$ LF with the Stepwise method.
This method assumes that the rest-frame luminosity function of
galaxies can be approximated by a binned distribution, where the
number density $\phi_i$ in each bin is a free parameter: this
non-parametric approach allows us to constrain the number density of
galaxies at different magnitudes without assuming an a priori
functional form (i.e. a Schechter-like shape).

We have assumed that the LF is made of four bins in the interval
$-22.5<M_{1500}<-17.0$, corresponding to the range sampled by our
observations. We also assume that galaxies are uniformly distributed
within the bins, with number densities $\phi{_i}$ to be determined.
With these simulated galaxies, we compute the
distribution of observed magnitudes originated from each bin, scaled
to the observed area in the WFC3 fields (subtracting the area covered by
detected objects). We then find the
combination of $\phi{_i}$ that best reproduces the magnitude
distribution of our observed objects with a simple $\chi^2$
minimization. Since the photometric scatter can move galaxies from a
magnitude bin to the near ones, the $\chi^2$ minimization ensures that
this effect is taken into account. The resulting uncertainties
can be higher than those derived assuming a simple linear relation
between the apparent and absolute magnitude, as in
\cite{bouwens08}. These differences are discussed in detail in
\cite{castellano10}.

Finally, the simulations are used to estimate the systematic
effects acting when we use colors at shorter wavelengths, i.e. in the
$BVI$ bands, to reject possible interlopers. Because
of the large IGM and internal
HI absorption, the expected flux in these bands for $z>6.5$ galaxies
is far below the detection threshold, or even null. For this reason a
stringent limit on the measured flux in these bands is adopted to
remove lower redshift interlopers. However, the S/N estimated by
SExtractor may be a poor representation of the actual photometric
scatter at low fluxes, due to a combination of factors, such as
uncertainties in the estimate of the local background, underestimates
of the true r.m.s., or chance superposition of faint blue galaxies
along the line of sight.
To account for these effects, we have measured the resulting
signal--to--noise (S/N) ratio in the $BVI$ images for each simulated object
inserted in the $J$ one, which should be zero on average. It turns out
that the actual distribution of the S/N ratios is wider than the one
obtained with SExtractor, which is computed scaling the input weight
image. We thus estimate the
``effective'' r.m.s. $\sigma_{S/N}$, i.e. the r.m.s. of the
signal--to--noise distribution in each of the $BVI$ images, which is
about 1.5 for the B and V bands and 1.8 for the I band.
Even taking into account this wider
distribution, we find that the tails of the S/N distribution
($S/N>2 \sigma_{S/N}$) contain more objects than in the case of a pure
Gaussian distribution.
As mentioned above, we will use the estimated $\sigma_{S/N}$ in all
BVI bands, requiring that high-z candidates have flux
$<2\sigma_{S/N}$ in all BVI bands and $<1\sigma_{S/N}$ in at least
two of them.  With our simulation, we estimate that the fraction of
true high-$z$ galaxies lost because of this strict criterion is about
$\sim12\%$ for the HUDF field. For the ERS, we find
an effective S/N ratio of 1.8 for all the BVI
bands and a rejection of true high-$z$ galaxies of $\sim24\%$.
Such an aggressive requirement
is needed in order to clean our sample by low-z
interlopers that are contaminating our z-dropout sample.
This technique has been extensively discussed in C10a and C10b.
In particular, if
we neglect the requirement that S/N must be $<1\sigma_{S/N}$ in at least
two bands
out of the BVI ones, we find 32 and 97 galaxy candidates for the HUDF and
ERS fields, respectively, to be compared with our 20 and 22 master
candidates. We thus prefer to reduce significantly
the numbers of possible contamination by interlopers, at the expenses of
being less complete by 24\%, especially for the ERS field.
This fraction is significantly higher than that found in the HUDF and it
is probably due to the fact that the depth of the ACS GOODS images
is not matched to the deep limiting magnitude one can reach with WFC3 in the
ERS. In this sense, these simulations are essential to estimate the LF at
$z\sim 7$.

\subsection{The morphological profiles used during simulations}

We expect that the output of the simulations depends critically on the
assumed morphology.
As realistic morphologies for the simulations related to the WFC3 fields,
we check different profiles. In particular, we explored the following
morphologies spanning from the most compact to the most extended objects
available in the WFC3 fields, exploring also plausible extrapolations from
lower-z studies:
\begin{itemize}
\item 1)
We select four point-like objects in the HUDF field based on their FWHM,
avoiding bright saturated objects. The spectroscopic classification is not
available for these objects, and their colors are consistent with those of
stars. Since these are compact, we use them as an input profile for very
compact galaxies at $z=7$, as it is expected that faint galaxies at high-z are
very small (\cite{windhorst02}).
\item 2)
The transient object at $Z-Y>3.6$ in Fig.\ref{fig:hudfcand} (probably a SNa)
is used as an empirical profile for our simulations.
Since this object was not detected in previous NICMOS images of HUDF
(\cite{oesch09}), it is plausible that the host galaxy in the background is not
able to change the SNa profile, and the result is a very compact object.
We used this template to check whether it gives different results
from the four stars described above.
\item 3)
We select the three brightest and most compact $z\sim 7$
candidates in the HUDF ($J<27.4$) as input
for our simulations, assuming that the physical dimension of high-z galaxies
does not vary dramatically with their magnitudes. We avoid in this group
the object HUDF-1095 (G2\_1408 of C10a) since it is the most extended
$z\sim 7$ candidate in our sample and we consider it in a separate group.
\item 4)
We study the effect of very extended objects or of merging at high-z using
as input in our simulations the object HUDF-1095 (G2\_1408 of C10a).
It is very bright with two knots, possibly the result of a merging of two
galaxies. Alternatively, its clumpy morphology
in the rest frame UV could be due to extended star forming knots in
a nascent galaxy. First results based on spectroscopic
confirmation with FORS2 seem to indicate that it is a $z=6.972$
LBG (\cite{fontana10}), confirming the photometric redshifts from ACS and
WFC3 based SED. Though not representative of the whole $z=7$ population, this
galaxy is an LBG at $z\sim 7$ and it is interesting
to study the resulting LF assuming this morphology as input for our
simulations.
\item 5)
There is the possibility that fainter galaxies are more compact than brighter
ones, as expected in the hierarchical formation models. To check this
hypothesis, we adopt as input profile the one obtained by stacking
the 15 fainter galaxy candidates at z=7 ($27.5<J<29.2$) in the HUDF field.
\item 6)
We used as input for our simulations the profiles of
spectroscopically confirmed galaxies at $z\sim 4$ in the ACS I-band
of the HUDF, scaled as $(1+z)^{-1}$ in their
physical sizes, as adopted in \cite{bouwens08} and \cite{bouwens10}.
Since there is
a regular size evolution with redshift, as found by \cite{ferguson04} and
confirmed by \cite{bouwens04} and \cite{oesch09b}, it is useful to explore
this case, in order to compare our results with earlier LFs that have
appeared in
the literature. These templates are also useful to avoid using
self-selected z=7 candidates as input for our simulations.
\end{itemize}

\begin{figure}[!h]
\includegraphics[width=9cm]{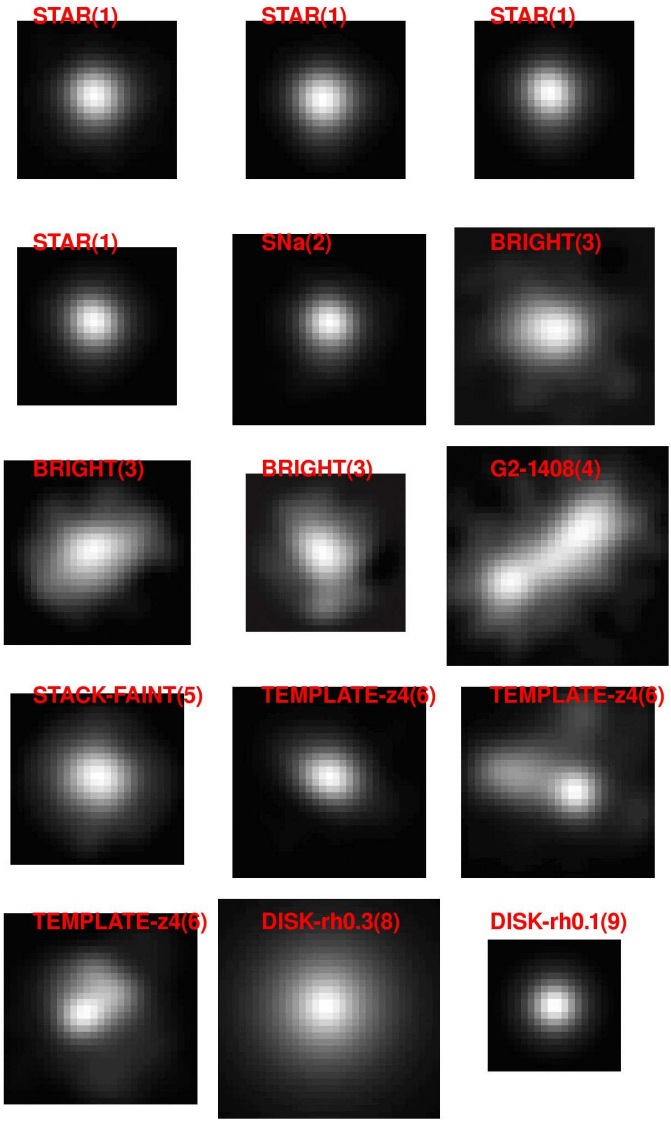}
\caption{
The different morphologies used for the simulations of WFC3 fields.
From top left, the first four cutouts represent non-saturated stars
in the HUDF survey (morphological template 1).
The SNa candidate outlined by \cite{oesch09} (template 2) is then shown,
followed by the three brightest $z\sim 7$
candidates of the HUDF survey (template 3).
The most extended galaxy is the brightest candidate in the HUDF
(HUDF-1095 or G2\_1408 of \cite{castellano09}, template 4).
The second thumbnail on the third raw shows the stack of the faint candidates
selected in the HUDF (template 5).
All these thumbnails have been extracted from the $J_{125}$ band of HUDF.
The remaining objects are spectroscopically confirmed galaxies
$z\sim 4$, extracted from the ACS-I band of the HUDF and shifted at z=7 as
described in the main text (template 6), and two artificial exponential
disk galaxies with half light radius=0.3 arcsec (templates 7 and 8) and
with half light radius=0.1 arcsec (templates 9 and 10).
}
\label{morphology}
\end{figure}

Fig.\ref{morphology} shows all the model galaxies
used for the HUDF and ERS simulations.

\begin{figure}
\includegraphics[width=9cm]{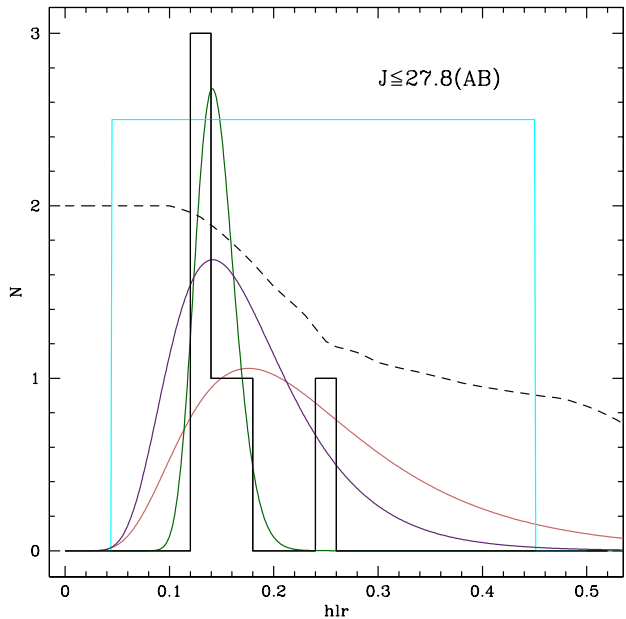}
\caption{
The black histogram shows the observed distribution of the half light radii
(in arcsec) of the $z\sim 7$ galaxy candidates brighter than $J=27.8$ in
the HUDF. At this magnitude limit the completeness of the J band image is
higher than 80\% even for the most extended objects. The galaxy with
half light radius $\sim 0.25$ arcsec is the HUDF-1095 one (template 4).
The cyan solid line shows
the uniform distribution in half light radius (template 7), while the
dark pink and green solid lines indicate the two log-normal
distributions in half light radius adopted by \cite{ferguson04}
(template 8, M1) and by \cite{oesch09b} (template 9, M2), respectively.
The purple line indicates the distribution derived from \cite{vanzella}
(template 10, M3). All sizes are corrected for PSF.
The dashed line shows the completeness of galaxy detection
for different half-light radii derived by the performed simulations.
This curve has been multiplied by a factor of 2
only for clarity.
}
\label{histrh}
\end{figure}

In addition of adopting single objects for the simulations, we explore also
different distributions of half light radii ($rh$) at $z\sim 7$:
\begin{itemize}
\item 7)
Following \cite{hudf05}, we explore a uniform distribution in the half
light radius of the
synthetic galaxies, assuming an exponential declining profile.
The two extremes of the uniform distribution are derived from the values
used by \cite{hudf05} for their simulations to recover the z=5 LF from HUDF
data. They adopt a uniform distribution
with half light radius from
0.05 to 0.5 arcsec. Converting these values to physical units at z=5, and
applying a $(1+z)^{-1}$ evolution until z=7, and reconverting in angular
dimensions, we derive the limits to be 0.045 to 0.45 arcsec.
Based on Fig.\ref{histrh}, it is clear that this distribution
can be ruled out by the present observations.
However, we decided to study this uniform distribution as
a ``thought experiment'' on how hypothetical size distributions might change
the derived completeness at faint magnitudes, but we do not use it
to estimate the $z\sim 7$ luminosity function.
\item 8)
Another possible distribution for the
half light radius of the simulated galaxies is the log-normal distribution
with mean and dispersion adopted following the relation of \cite{ferguson04}
for photometrically selected LBGs at z=4,
applying a $(1+z)^{-1}$ evolution to the half light radii in physical units.
We derive for the z=7 population a mean half light radius of 0.258 arcsec and
$\sigma=0.0358$.
This approach has been adopted also in \cite{hudf09} and \cite{oesch09}.
In the following it is indicated also as M1.
\item 9)
The typical half light radii of z=7 candidates are of the order of
0.1-0.15 arcsec. As in \cite{oesch09b}, we assume a log-normal distribution
of the half light radius of the simulated galaxies with mean=0.153 and
sigma=0.0574 arcsec. This distribution, however, cannot explain the
presence of the HUDF-1095 galaxy, with $rh>0.25 arcsec$.
In the following it is indicated also as M2.
\item 10)
A different distribution for the
half light radius of the simulated galaxies at $z\sim 7$ has been derived
starting from the observed half light radii of spectroscopically confirmed
LBGs at $z=4$ in GOODS from \cite{vanzella}. We select only galaxies with
$24\le I\le 25$ magnitudes to avoid incompleteness for extended LBGs.
Applying a $(1+z)^{-1}$ evolution to the half light radii at $z\sim 4$
in physical units, the log-normal distribution has a mean=0.177 and
sigma=0.0051 arcsec at $z\sim 7$ and in apparent dimensions.
In the following it is indicated also as M3.
\end{itemize}

From Fig.\ref{histrh} it is clear that the observed
distribution of half light radii for the z=7 candidates peaks at $\simeq 0.1$
arcseconds and extends to 0.2-0.25 arcsec. The small area covered
by the two WFC3 surveys considered here is at present too small to
sample the extended tail of the half light radii distribution, and it is not
possible to infer a robust size distribution.
Clearly, the uniform distribution (criterion 7) is not representative of
the observed $z=7$ galaxy population and will not be considered
in the following to estimate a reliable $z\sim 7$ luminosity function.

\begin{figure}
\includegraphics[width=9cm]{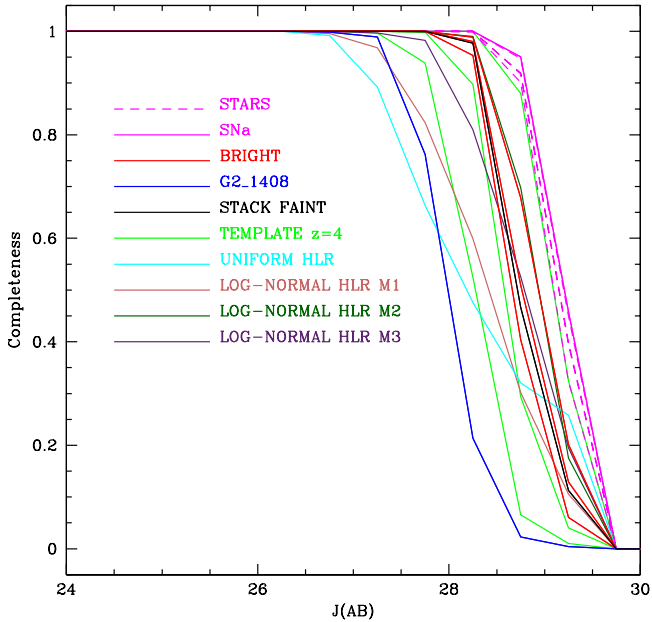}
\caption{
The fraction of detected galaxies versus the $J_{125}$
input magnitude for the simulations in the HUDF field.
Different colors indicate the various morphologies adopted:
the dashed magenta lines indicate non-saturated stars
on the HUDF survey (morphological template 1). The solid magenta line shows
the completeness obtained using the SNa candidate
outlined by \cite{oesch09} (template 2) as morphological template,
while the red solid curves show the three brightest $z\sim 7$
candidates of the HUDF survey (template 3).
The blue solid line is the completeness derived by adopting
as template the brightest candidate in the HUDF (HUDF-1095 or G2\_1408 of
\cite{castellano09}, template 4), and the black solid line
represents the completeness obtained adopting as input template
the stack of the faint candidates selected in the HUDF (template 5).
The green dashed lines indicate the spectroscopically confirmed galaxies
$z\sim 4$, shifted at $z=7$ (template 6).
The cyan solid line shows the completeness obtained assuming
a uniform distribution in half light radius (template 7), while the
dark pink and green solid lines indicates the results of two log-normal
distributions in half light radius adopted by \cite{ferguson04,hudf05,oesch09}
(template 8, M1) and by \cite{oesch09b} (template 9, M2), respectively.
The purple line indicates the completeness derived from \cite{vanzella}
(template 10, M3).
}
\label{completeness}
\end{figure}

The three log-normal distributions (templates 8, 9, and 10
shown in Fig.\ref{histrh})
cannot be confirmed or ruled out at the present stage for a number of reasons:
\begin{itemize}
\item a)
the $z\sim 7$ candidate sample at magnitude $J<27.8$ in the HUDF
is limited to 6
galaxies. We have chosen this magnitude limit since it is known from
simulations that in the HUDF we are complete at the $\sim 80\%$ level for
extended galaxies (see Fig.\ref{completeness}, blue and cyan lines).
The corresponding limits for shallower surveys (ERS or the parallel HUDF
observations HUDF-P1 and HUDF-P2) are 1-2 magnitude brighter
($J<26.0-26.8$) than in the HUDF.
Adopting these limits, only two additional brighter candidates
have been found on the ERS survey, both having $rh>0.2 arcsec$.
Since the number
statistic here is very low due to the limited area covered by ultradeep HST
observations, at the present stage we cannot exclude the presence of
more extended galaxies at $J<27.8$ magnitude.
\item b)
the extrapolation to fainter magnitudes of the size distribution
for $z\sim 7$ galaxies is not known at present.
It is possible that
a population of extended sources (similar to HUDF-1095) exists at high-z
and at fainter magnitudes. The present HUDF observations cannot rule out such
hypothesis.
\item c)
the spectroscopic confirmation of $z\sim 7$ galaxy candidates
with detailed HST morphology and half light radius determination
in the HUDF and ERS fields is
still lacking at this stage and some of them
could be lower-z contaminants. The brightest galaxy in the HUDF
(HUDF-1095, which is clumpy or in a merging state) turned out to
be spectroscopically confirmed at z=6.972 (\cite{fontana10}).
Since it is the only galaxy with detailed HST morphology in the NUV
rest frame (template 4) to be confirmed at $z\sim 7$ in the HUDF field,
and it is relatively bright ($J=26.07$) and extended ($rh=0.256$),
we cannot exclude that a similar population of extended sources exists at
high-z and at fainter magnitudes.
\item d)
in \cite{bouwens10c}, and also in \cite{wilkins2},
a $z\sim 7$ galaxy candidate with an half-light radius of
0.46 arcsec has been found (UDF092z-01191133). Moreover, other $z\sim 7$
candidates
found in the ERS, HUDF-P1, and HUDF-P2 fields show extended half light
radii (0.3 arcsec or more). If they turn out to be spectroscopically
confirmed at $z\sim 7$, then distributions with extended half light radii,
similar to template 8 (M1), would become more reliable.
\end{itemize}

Based on these considerations, it is not possible at the present stage
to exclude any of the adopted log-normal half light radii
distributions (M1, M2 or M3).

The simulations carried out with the different morphological templates
described above have been
used to estimate the completeness in the detection procedure. This is
shown in Figure~\ref{completeness}, where we plot the
fraction of detected versus input objects as a function of the $J_{125}$
input magnitude for the HUDF field.

The completeness for the HUDF is basically equal to 100\% down to
$J\simeq 27$ for all the adopted morphologies, and fades in different
ways according to the different half light radii of the templates
adopted in the simulation. In particular, more extended galaxies typically
have a lower completeness at $J>27$, and {\em the fraction of faint z=7
galaxies recovered depends dramatically on the morphological templates
used during the simulations}.
Since the completeness is used to estimate the LF,
we expect the faint end of the z=7 LF to be strongly dependent on
the morphological template adopted, as we will show in the next section.

%__________________________________________________________________

\section{Results}

The derived LF is sensitive to the
details of the simulations adopted. Here we are interested in
exploring mainly two issues: the role of the color selection criteria
on the estimate of the LF and the role of galaxy morphology on the
measurement of $\alpha$, the steepness of the faint end of the LF.

\subsection{Results for different color selections}

\begin{figure}
\includegraphics[width=9cm]{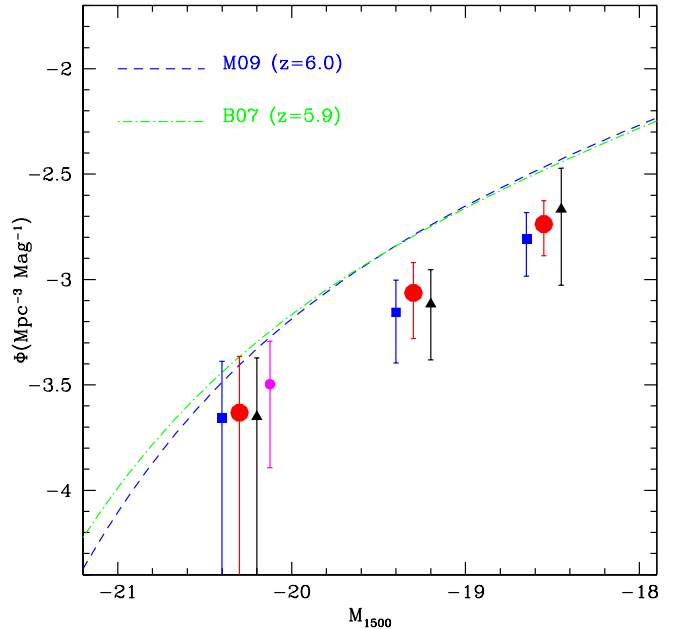}
\caption{
The LF at z=7, obtained using as morphological templates real bright
($J<27.4$) candidates at $z\sim 7$ found in the HUDF (template 3). The magenta
circle represents the LF found in the Hawk-I from C10a,
while the other
symbols indicate the LF in the HUDF and ERS derived
adopting different color criteria: our color criteria (red filled
circles, criterion C), \cite{bunker09} (small filled black triangles,
criterion B) and \cite{oesch09} (blue squares, criterion A).
The blue points are shifted by -0.1 magnitudes and the black points by +0.1
for clarity.
The blue dashed and green dot-dashed lines show the LF at z=6 of
\cite{mclure09} and \cite{bouwens07}, respectively.
}
\label{LFZ7obs}
\end{figure}

We first explore three different color criteria for the selection of
$z\sim 7$ galaxies. The LF found with these three selection criteria (A, B,
and C), described in Section 3, for the HUDF and ERS fields is shown in
Fig.\ref{LFZ7obs}. The red filled circles show the LF derived with the color
criterion C, compared to the one derived with the B (small
filled black triangles) and A criteria (blue squares), for
the HUDF data alone. We plot for comparison also the Hawk-I results
(magenta open circle) of C10a. The WFC3 LF has been
derived assuming as morphological inputs for our simulations
real bright ($J<27.5$)
candidates at $z\sim 7$ found in the HUDF (template 3).
Adopting as morphological templates
spectroscopically confirmed galaxies at $z\sim 4$ found in the ACS
i-band of the HUDF and scaled to $z\sim 7$ (template 6),
or the stack of the fainter ($J>27.5$) candidates (template 5),
we obtained very similar results.
We have also verified that the additional limit
$J_{125}-H_{160}\ge -1.0$ adopted in criterion C does not
imply significant changes in the LF in the faintest bin ($M_{UV}\sim -18.5$).

It is clear that the three color selection criteria analysed here
give comparable results, well within the uncertainties, despite the size
of the three samples changes by a factor of $\sim$3. This is due to
the fact that all the three criteria are well designed to isolate the
few $z\sim 7$ candidates from the great number of interlopers, but each
method makes different assumptions, resulting in a variable number of
z-dropout galaxies. {\em The resulting LFs, however, are similar, since the
simulations required to derive them are able to take into account properly
the random and systematics effects (incompleteness and contamination
by photometric scatter), especially at the faintest
magnitudes}. In the faintest magnitude bin the sizes of the
error bars depend on the criteria adopted, with the larger errors for the
most strict criterion. The B criterion selects indeed a smaller number
of candidates,
with a resulting higher Poisson noise than the one derived adopting the
A or C criteria.

The only difference between the A and C criteria is
the magnitude cut adopted: we select all galaxies down to $J=29.2$, while
\cite{oesch09} recover only galaxies with $S/N$ greater than 5 in the Y and J
bands. These color criteria are similar, and no systematic differences are
found in the LF estimate. For this reason, we decide to adopt the C color
criterion in the following sections.

\subsection{Results for different morphologies}

\begin{figure}
\includegraphics[width=9cm]{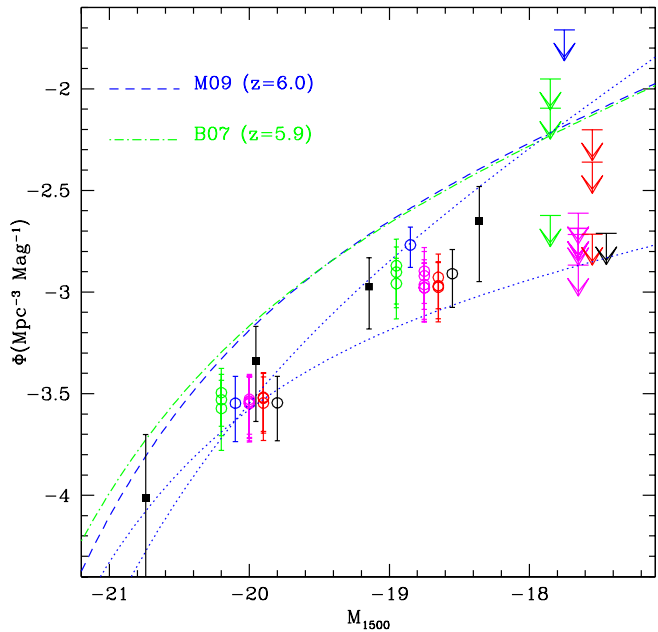}
\caption{
The LF at z=7, obtained using as morphological types galaxies drawn from
the templates 1-6. The color code is the same of Fig.4.
The blue and green bins are shifted by -0.2 and -0.1
magnitudes, while the red and black points are moved by +0.1 and +0.2,
respectively, for clarity.
The blue dotted lines indicates the range of variability
for the best fit of $\alpha$ spanning the range $-2.15\le \alpha\le -1.38$.
The blue dashed and green dot-dashed lines show the LF at z=6 of
\cite{mclure09} and \cite{bouwens07}, respectively.
The filled black squares are the LF presented in \cite{oesch09}.
}
\label{fig:LFZ7morph}
\end{figure}

We then explore the issue of adopting different morphological templates
in the simulations.
For each morphological type shown in Fig.\ref{morphology} and discussed
in Section 4, we have simulated
$2\times 10^4$ synthetic galaxies as described in Sect. 4 and used them
to compute the LF at $z\sim 7$.
In Fig.\ref{fig:LFZ7morph} we show the LF for each template considered:
at $M_{1500}\sim -20$, the different morphologies adopted have a small
impact on the number density $\Phi$, while at $M_{1500}\sim
-18$ the LF is extremely dependent on the assumption of the morphological
templates. In Fig.\ref{fig:LFZ7morph} the open points show the result of
templates 1-6, based on single object morphologies, with the relevant
uncertainties.

\begin{figure}
\includegraphics[width=9cm]{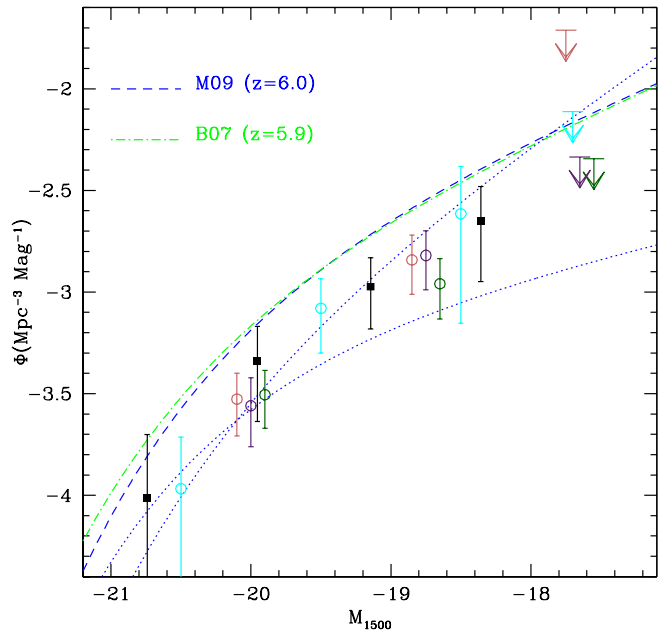}
\caption{
The LF at z=7, obtained using different distributions for
the simulated half light radii, namely the uniform one (template 7,
cyan points), and the three log-normal distributions (template 8 using dark
pink points, template 9 using dark green points, and template 10 using
purple points).
The dark green bins are shifted by +0.1 magnitudes, while the dark pink
points are moved by -0.1 magnitudes,
respectively, for clarity.
The blue dotted lines indicate the range of variability
for the best fit of $\alpha$ spanning the range $-2.15\le \alpha\le -1.38$.
The blue dashed and green dot-dashed lines show the LF at z=6 of
\cite{mclure09} and \cite{bouwens07}, respectively.
The filled black squares are the LF presented in \cite{oesch09}.
}
\label{fig:LFZ7morphDIST}
\end{figure}

In Fig.\ref{fig:LFZ7morphDIST} we considered different distributions for
the simulated half light radii, namely the uniform one (template 7,
cyan points), and the three log-normal distributions (template 8 using dark
pink points, template 9 using dark green points, and template 10 using
purple points). For comparison, we plot the LF derived by \cite{oesch09} as
filled black squares: these are in agreement
with the result based on the simulations carried out with the distributions
of half light radii which extend to 0.2-0.4 arcsec.
The large variance in the galaxy number density at the faint end
of the $z\sim 7$ LF depends on the different completeness corrections
applied to the same dataset and thus reflects the uncertainties due
to the poorly known half light radius distribution.

Fixing the Schechter parameters $\Phi^\ast$ and $M^\ast$ to
$6.9\times 10^{-4}$ and -20.10 (the best fit values of \cite{ouchi}),
respectively and varying $\alpha$,
we derive a large range of variation for the best fit of
the LF steepness, namely  $-2.15\le \alpha\le -1.38$ at 68\% c.l.,
represented by the two blue dotted lines in the plot.

\begin{figure}
\includegraphics[width=9cm]{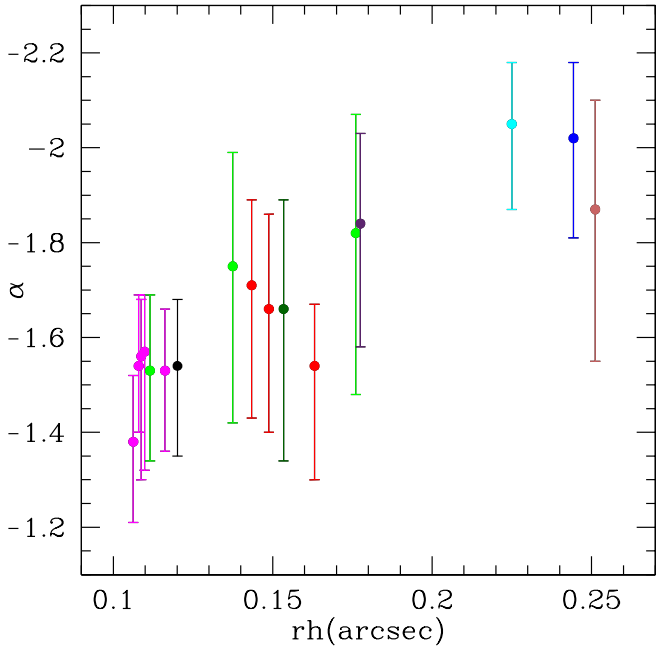}
\caption{
The dependence of the steepness of the LF, $\alpha$, against the
half light radii of the simulated galaxies.
Different colors indicate various morphology adopted, using the same
color code of Fig.4.
}
\label{alpharh}
\end{figure}

The steepness of the faint end of the LF, $\alpha$, depends critically
on the half light radii of the synthetic galaxies used to carry out
the simulations. In Fig.\ref{alpharh} we show the half light radii
of the different morphological templates against the best fit $\alpha$
of the LF. For large value of the half light radius, namely $>0.2$
arcsec, the best fit $\alpha$ is between -1.8 and -1.9, as found by
\cite{oesch09}, \cite{bouwens10}, and \cite{bouwens10c} at z=7 and beyond.
For smaller values, $\sim 0.1$ arcsec, we find
$\alpha$ in the range -1.4 to -1.7, in marginal
agreement with \cite{mclure10}
that have used point sources as input for their simulations.
The uncertainties on this
parameter are still large, of the order of $\sim 0.2-0.4$. {\it We
find however a significant relation between the half light radius and
$\alpha$, indicating that simple conclusions based on the LF at $z\sim
7$ depend dramatically on the half light radius distribution assumed
for the simulations.}

\subsection{Fitting the LF at $z\sim 7$}

We want to derive here the LF at $z\sim 7$ making reliable and robust
assumptions based on the considerations outlined above. We adopt the
color criterion C, with a magnitude cut at J=27.4 and J=29.2 for the
ERS and the HUDF, respectively. As shown above, the simulations are
able to correct the incompleteness of this color criterion in a
reliable way. The morphology templates used in the simulations,
however, are critical, since the simulations cannot correct for a
biased assumption on the profiles of the candidates.
Since it is not possible, with the small number statistics achieved by
the present WFC3 data, to derive precisely the distribution
of the half light radii of the $z\sim 7$ LBGs, especially at the faint end and
at high values of $rh$, we carry out the fit for
the two most extreme, but reliable, variants of the LFs derived
in the previous paragraph.
We have adopted the simulations carried out with the stacking of the faint
candidates selected in the J band of HUDF (template 5) as an extreme
case (all the z=7 galaxies are compact), while the HUDF-1095 galaxy
(template 4) has been chosen as the opposite case (all high-z LBGs are
very extended). We have verified that all the other options (templates 1-3
and 6-10) give results that are in between these two extremes.

\begin{figure}
\includegraphics[width=9cm]{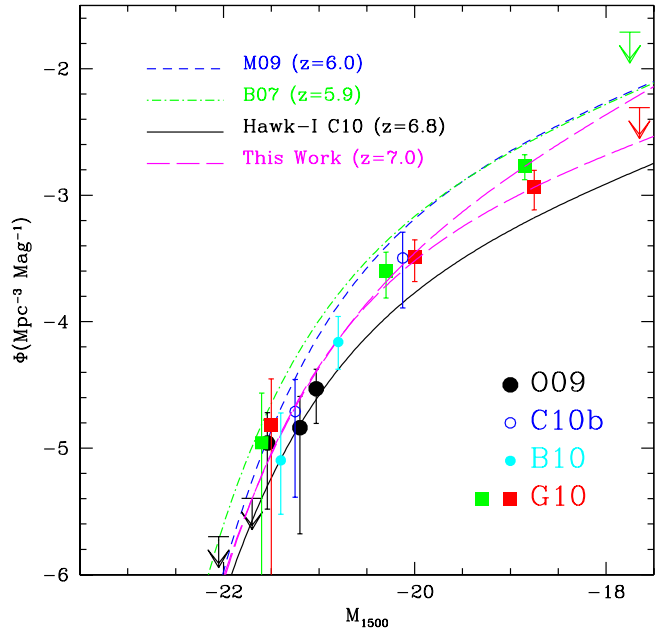}
\caption{
The LF at z=7, obtained using as morphological templates the stacking of the
faint candidates at $z\sim 7$ found in the J band image of the HUDF
(template 5) and the C color criterion. The LF is computed using
ERS and HUDF fields together (red filled squares, G10).
The green filled squares show the LF obtained with the same color
criterion C but using the HUDF-1095 candidate as input for the
simulations (template 4).
These data have been complemented with the results of \cite{ouchi}
(O09), \cite{bouwens10nic} (B10) and C10b.
The lower dashed magenta line
is the best fit to all the observed points related to the template 5,
the upper dashed magenta line is the best fit
related to the template 4, while the black solid line
shows the Hawk-I result of C10a.
The blue dashed and green dot-dashed lines show the LF at z=6 of
\cite{mclure09} (M09) and \cite{bouwens07} (B07), respectively.
}
\label{LFZ7new}
\end{figure}

We have combined the results of the ERS and HUDF
fields in a single LF, to reduce the statistical
uncertainties and to have an extended range in the absolute magnitude
$M_{1500}$. To cover the bright part of the LF, we also use the
results of \cite{ouchi} obtained from a shallow and wide Subaru survey
and those of C10a and C10b derived from an intermediate
depth survey with Hawk-I.
In Fig.\ref{LFZ7new} we show our results (red and green filled
squares) complemented with the results of \cite{ouchi} (filled black
circles), \cite{bouwens10} (filled cyan circles) and
of C10a and C10b (empty black circles).

\begin{figure*}
\includegraphics[width=9cm]{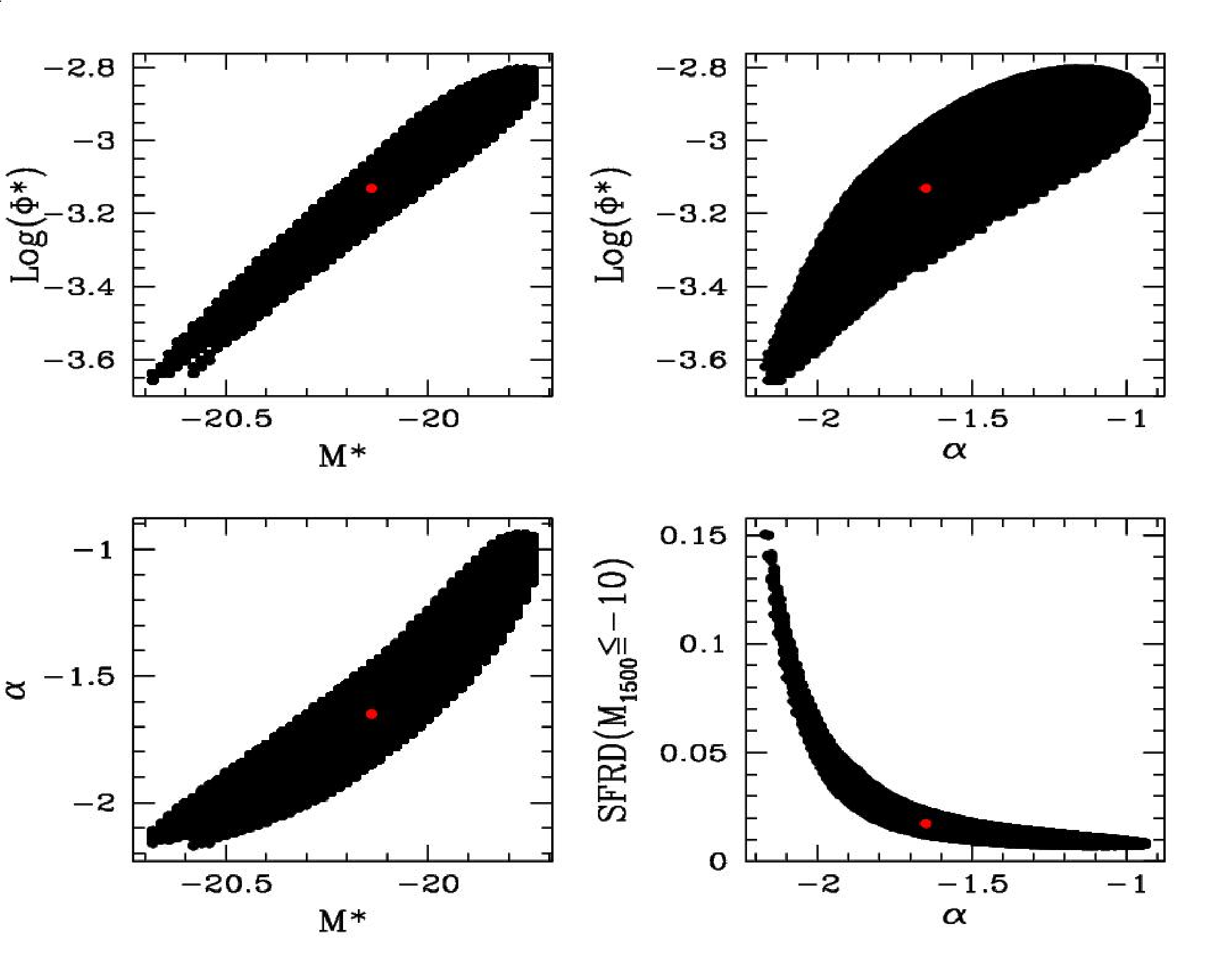}
\includegraphics[width=9cm]{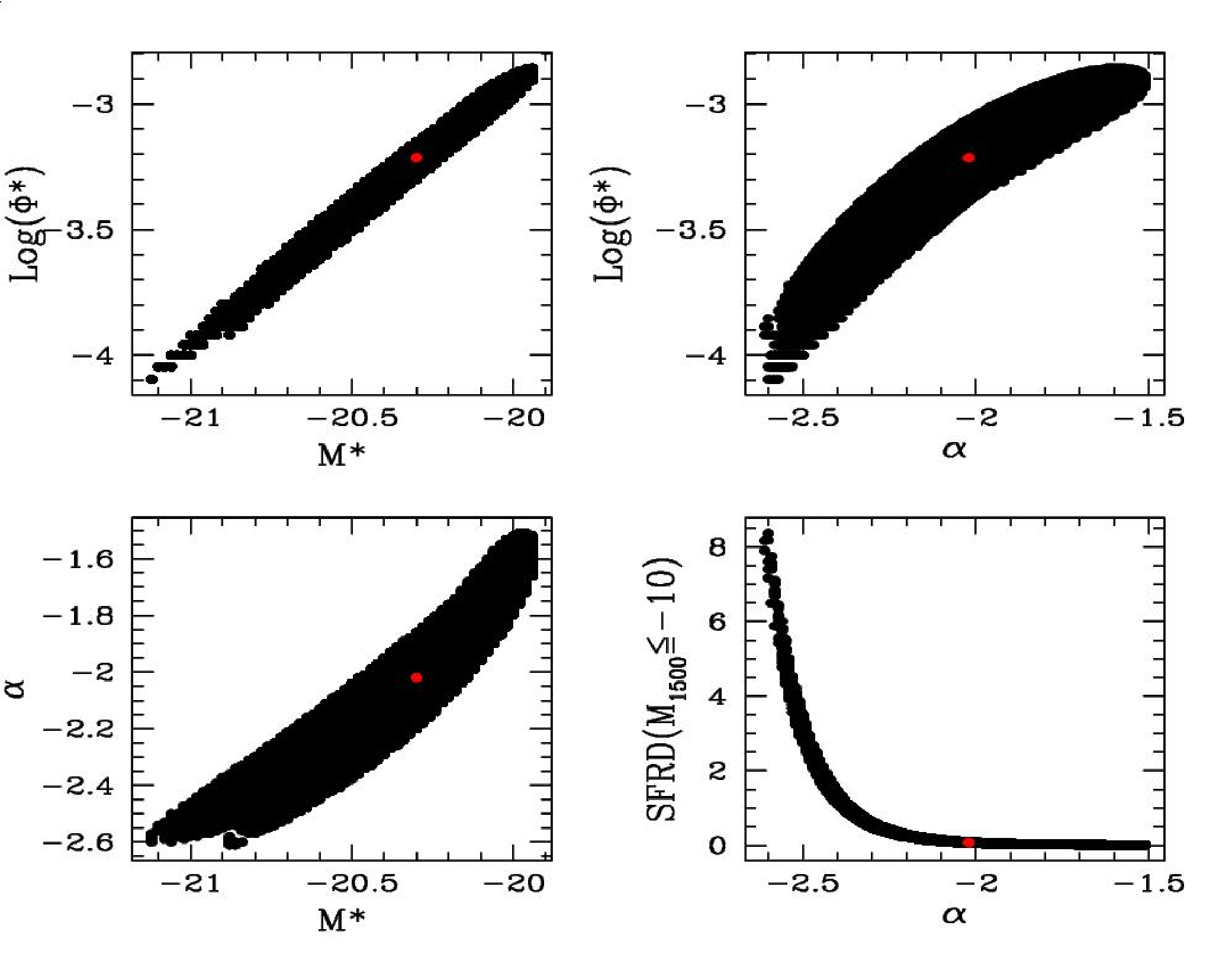}
\caption{
The confidence regions at 68\% c.l. (1 $\sigma$) for
the LF parameters and the SFRD. The red points show the projection of the
best fit.
The plot on the left shows the confidence regions based on the simulations
with compact morphology (template 5), while the right plot is related
to extended morphology (template 4).
}
\label{fig:fitparam}
\end{figure*}

All the observed points shown in Fig.\ref{LFZ7new} have been used
to fit the LF. We scan the parameter space in 3D ($\Phi^{\ast}$, $M^{\ast}$,
and $\alpha$) adopting a Schechter parameterization (\cite{Schechter1976})
and taking into account
the asymmetric errors and the presence of upper limits. The best fit
is shown (long dashed magenta lines), and it is compared to the early Hawk-I
result of C10a (black solid line) and the LFs at z=6 of
\cite{bouwens07} and
\cite{mclure09}. Using our best fit, we provide a robust
estimate of the LF at $z=7$, excluding at $>$99.73\% c.l.
($>3\sigma$) the non-evolution from z=6 to z=7. Indeed, we are able to show
that the
normalization at $M_{1500}\sim -19$ is significantly lower than the LF at
$z\sim 6$. Moreover, combining data with a
large range in absolute magnitudes we are able to reduce the degeneracies
in the three parameters of the LF. In Fig.\ref{fig:fitparam} we provide
the confidence regions at 68\% c.l. for $\Phi^{\ast}$, $M^{\ast}$,
and $\alpha$ at z=7, exploring also the constraints given on the
SFR density (SFRD).

%__________________________________________________ One column table
\begin{table}
\caption{Schechter LF fit Results}
\label{table:lffit}
\centering
\begin{tabular}{c | c c c}
\hline\hline
Compact Morphology & Best Fit & Lower(1$\sigma$) & Upper(1$\sigma$) \\
\hline
$Log(\Phi^\ast)$ & -3.13 & -3.66 & -2.80 \\
$M^\ast$ & -20.14 & -20.68 & -19.74 \\
$\alpha$ & -1.65 & -2.17 & -0.94 \\
$\rho_{UV}(-10)$ & 9.2E+25 & 3.8E+25 & 7.9E+26 \\
$SFRD(-10)$ & 0.017 & 0.007 & 0.150 \\
$\rho_{UV}(-19)$ & 2.7E+25 & 1.9E+25 & 3.5E+25 \\
$SFRD(-19)$ & 0.005 & 0.004 & 0.007 \\
\hline
\hline
Extended Morphology & Best Fit & Lower(1$\sigma$) & Upper(1$\sigma$) \\
\hline
$Log(\Phi^\ast)$ & -3.22 & -4.10 & -2.86 \\
$M^\ast$ & -20.30 & -21.12 & -19.94 \\
$\alpha$ & -2.02 & -2.61 & -1.51 \\
$\rho_{UV}(-10)$ & 4.5E+26 & 8.8E+25 & 4.4E+28 \\
$SFRD(-10)$ & 0.085 & 0.017 & 8.365 \\
$\rho_{UV}(-19)$ & 3.2E+25 & 2.3E+25 & 3.9E+25 \\
$SFRD(-19)$ & 0.006 & 0.004 & 0.007 \\
\hline
\end{tabular}
$\rho_{UV}(-10)$ is the luminosity density obtained integrated the LF down
to an absolute magnitude of -10 based on a simple extrapolation, while
$\rho_{UV}(-19)$ is the
one derived integrating the best fit LF down to an absolute magnitude
$M_{1500}$ of -19. The $SFRD(-10)$ and $SFRD(-19)$ represent the SFR density
limited to -10 and -19 in $M_{1500}$. The unity is in $erg/s/Hz/Mpc^3$ for
$\rho_{UV}$ and in $M_\odot/yr/Mpc^3$ for SFRD.
The upper part of the table summarizes the LF parameters resulting from
simulations
based on template 5 (compact morphology), while the lower part is related
to template 4 (extended morphology).
\end{table}

Table \ref{table:lffit} summarizes the best fit parameters for the
Schechter LFs, with their 68\% c.l. uncertainties. The luminosity
density $\rho_{UV}$ has been computed integrating the fitted LFs down
to an absolute magnitude (at 1500 \AA~ rest frame) of $M_{UV}=-10$
and to $M_{UV}=-19$.
We have converted these values in a SFRD
following the standard formula by Madau et al. (1998) and applying the
extinction correction of Meurer et al. (1999) of $A_{1600}=0.43$
(considering an average
UV slope $\beta=-2.0$ for the SED of the z=7 LBG candidates).
Our estimates of the luminosity density and SFRD are consistent with
other values in the literature, reducing
significantly their uncertainties.

%__________________________________________________________________

\section{Discussion}

We discuss the implications on reionization derived from our best fit
of the LF and the uncertainties on the LF parameters implied by the
assumption made during completeness simulations.

\subsection{Reionization}

\begin{figure}
\includegraphics[width=9cm]{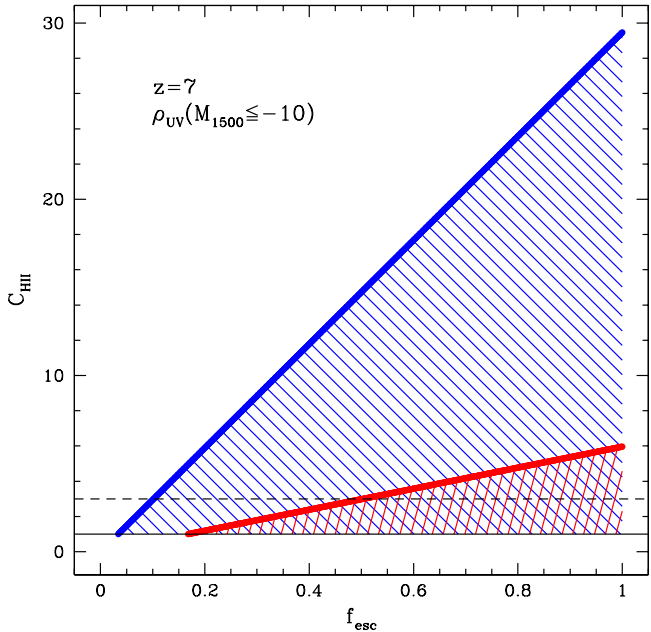}
\caption{
The constraints on the maximum $C_{HII}$ and minimum $f_{esc}$ parameters
in order to have the Universe reionized by LBGs at z=7 given our LF.
Given a value for $f_{esc}$, the Universe is reionized if $C_{HII}$ is
lower than the red line, if a compact morphology is used (template 5),
or lower than the blue line if an extended one is considered (template 4).
The $\rho_{UV}$ is computed integrating the LF down to
$M_{1500}=-10$ magnitude.
The solid line at $C_{HII}=1$ indicates a homogeneous IGM, and we
must consider only the $C_{HII}\ge 1$ region, since galaxies at $z\sim 7$
are formed in biased density regions of the Universe. The dashed line shows
the value $C_{HII}=3$.
In practice, only a combination of ($f_{esc}$,$C_{HII}$) parameters inside
the shaded regions corresponds to a $z\sim 7$ IGM kept ionized by LBGs
only.
}
\label{reion}
\end{figure}

Following \cite{Bolton2007} and \cite{ouchi},
the contribution of z=7 LBGs to the hydrogen ionizing photon can be set as

\begin{equation}
\dot{N}_{ion}(s^{-1}Mpc^{-3})=10^{49.7}(\frac{\rho_{UV}}{6\cdot 10^{25}})
(\frac{3.0}{\alpha_S})(\frac{f_{esc}}{0.1})
\end{equation}

where $\alpha_S$ is the spectral index of ionizing emission and
$f_{esc}$ is the escape fraction of ionizing photons. The rate of
ionizing photons needed to balance the recombination process of
hydrogen in the IGM and hence to keep the Universe reionized is
$\dot{N}_{ion}\ge 10^{47.4}C_{HII}(1+z)^3$, that translates into a
constraint to the clumpiness of the IGM to be $C_{HII}\le (\rho_{UV}
f_{esc})/(10^{22}\alpha_S (1+z)^3)$. We fix these parameters to
$z=7$, $\beta=-2$, and $\alpha_S=3$ in the following.
In Fig.\ref{reion} we show in the dashed areas the $C_{HII}$ allowed by
a given $f_{esc}$ to keep the IGM ionized at $z\sim 7$: the red region is
for a compact morphology (template 5), while the blue one refers
to an extended morphology (template 4). The relevant UV emissivity of LBGs
$\rho_{UV}$ is computed by
integrating the LF down to $M_{1500}=-10$ (\cite{bouwens10c}),
assuming that the steepness of
the faint end of the LF remains constant; we have no information on the
number density of $z\sim 7$ galaxies at these faint magnitudes with
the present data. Given our confidence intervals for the LF
parameters, and hence the limit to $\rho_{UV}(-10)$, we
derive a limit $C_{HII}\le 6.5$ (at 68\% c.l.) in order to have the
Universe ionized at $z=7$, assuming a maximum escape fraction of 1.0,
and the simulations based on template 5 (compact morphology),
as shown in Fig.\ref{reion}. If we use the LF based on
extended morphology (template 4), the 68\% c.l. limit to the clumpiness
of the ionized hydrogen is $C_{HII}\le 29.6$.
In this reasoning we must consider
$C_{HII}\ge 1$, since galaxies at $z\sim 7$ are formed in biased
density regions of the Universe and the IGM is plausibly not
homogeneous ($C_{HII}=1$) at these redshifts. Limiting the clumpiness
to $C_{HII}\ge 1$, we thus have $f_{esc}\ge 0.17$ in order to have the
Universe reionized by z=7 assuming compact morphology,
while the limit is $f_{esc}\ge 0.034$ for extended morphology.

Bolton \& Haehnelt (2007) have inferred $C_{HII}\le 3$ at $z\sim 6$ from the
Lyman-$\alpha$ forest photoionization state: since $C_{HII}$ is expected to
monotonically decrease towards high-z in a hierarchical Universe,
an escape fraction of $f_{esc}\sim 0.5$
is enough for the reionization of the IGM (but not for the more clumpy
galaxy formation regions) considering template 5 (compact morphology),
while a value of $f_{esc}\sim 0.1$ is sufficient according to template 4
(extended morphology).

We neglect in our computation the contribution of AGNs, since their
LF at z=7 is still unknown and the upper limits currently available
indicate that the AGN will add only 5-8\% to the luminosity density of
galaxies (\cite{ouchi}).

The reionization rate $\dot{N}_{ion}$ however depends largely on the
contribution of faint galaxies ($-19\le M_{1500}\le -10$), since
$\rho_{UV}(M_{1500}<-10)$ could be 10-15 times larger than
$\rho_{UV}(M_{1500}<-19)$ if the LF is steeper than $\alpha \le -2$.
This is eventually still possible, if a large number of faint and extended
galaxies exist, similar to the object HUDF-1095
(G2\_1408 of \cite{castellano09}).
If the faint end of the LF at z=7 is not so steep, as we have inferred here
for the compact morphologies,
the contribution of fainter galaxies ($-19\le M_{1500}\le -10$) is only
2.4 times that of $M_{1500}\le -19$ LBGs.
{\em Thus, the present uncertainty on the steepness of the faint end of the LF
translates in a large uncertainty, of
a factor 2-15, on the best fit of the luminosity density, SFR density,
and reionization rate at z=7.}

Following \cite{windhorst02}, it is plausible that fainter galaxies have 
smaller physical sizes: in this case a combination of high escape fraction
($f_{esc}\ge 0.2$) and small clumpiness ($C_{HII}\le 3$) is required for
the reionization of the Universe. It is worth noticing that the recent
estimates of $f_{esc}$ for $L\ge L^\ast$ LBGs at $0\le z\le 3$ are
$\le 0.15$ (\cite{bridge10,cowie10,siana10,vanzella10}), implying
a consistent evolution of the galaxy escape fraction going to fainter
luminosities or to higher redshifts.

Moreover, the knowledge of the reionization processes in the high-z Universe
is currently far from clear. For example,
early theoretical works (i.e. \cite{gnedin}) have found high values
for the clumpiness, $C_{HII}\sim 10-30$.
More recently, \cite{pawlik09} studied the
effect of photoionization heating by a uniform ultraviolet background
and found that photo-heating can strongly reduce the clumping factor of the
IGM because the increased pressure support smoothes out small-scale
density fluctuations. The common expectation is that
photoionization heating should provide a negative feedback
on the re-ionization of the IGM because it suppresses the cosmic SFR
by boiling the gas out of low-mass DM halos. However, since it also contribute
to the reduction of the clumping factor, the net result is that it is easier
to keep the IGM ionized. Photo-heating therefore also provides a positive
feedback for the reionization processes. Neglecting the effects of
photoionization heating, \cite{pawlik09} predict a clumpiness of the IGM of
the order of 7-8 at z=7, and in this case our limit to the LF rule out a
completely ionized IGM even for $f_{esc}=1$ if all the galaxies at $z=7$
resemble our morphological template 5 (compact galaxies), while an
$f_{esc}\sim 0.3$ is still sufficient assuming template 4 (extended galaxies).
If instead the photoionization
heating is considered, a small clumpiness $C_{HII}\le 3$ is expected,
in agreement with measurements by Bolton \& Haehnelt (2007). In this case,
an escape fraction of $\sim$50\% is enough to keep the Universe re-ionized,
even in the most pessimistic case (of compact morphology).

\subsection{The reliability of our LF estimate}

The uncertainties on the estimate of the $z=7$ LF are related to the method
used to derive it. Errors can be due to the limited
depth of the images used, the color criteria adopted,
the presence of possible interlopers (cool dwarf stars, low-z red galaxies,
transient objects), spurious
detections, especially when going faint. All these features have been
discussed in the previous sections, showing that detailed simulations are
able to take into account the effects of different color criteria, and
that a precise color cut can be identified with the aim of separating
the few high-z galaxies from the much frequent interlopers.
Moreover, the details on the steepness of the faint end of the LF are strongly
influenced by the morphological templates adopted for the simulations.

There are however two issues that cannot be corrected with the
simulations only: the presence of transient objects within the
z-dropout candidates and the role of the cosmic variance when dealing
with deep, pencil beam surveys. In the near future e.g., the HUDF
field will be re-observed with WFC3 in order to go deeper than the
present observations, with 34 additional orbits splitted
in three filters (Y, J, and H), complementing the existing 60 orbits
discussed here. With these data, it will be possible to assemble a
z-dropout sample clean from variable contaminants. The
discovery of a clear transient source in the HUDF data (see
Fig.\ref{fig:hudfcand}) and two possible transients in the Hawk-I data
by \cite{castellano09} is more than an order of magnitude higher than
the 0.06 SNe per WFC3 field estimated in \cite{oesch09}. It is possible that
some of the selected z=7 candidates will be rejected as interlopers
when a second epoch survey will be executed.

Surveys are usually bounded by a limited amount of telescope time, and thus
they have to compute a trade off between the covered area and the imaging
depth. This has deep implications on the number of objects recovered, on
their absolute magnitudes and the sampled Universe volume.
Simple observational estimates of galaxy number counts in finite
volumes are thus subject to the uncertainties due to cosmic variance,
arising from underlying large-scale density fluctuations.
This may be particularly true at high-z, where galaxies are expected to
be much more clustered than at lower-z.

We have used the Millennium simulation, using the realization of \cite{kw},
to investigate the impact of the cosmic variance on the number of galaxies
recovered in the two WFC3 fields analysed here. We extract randomly from the
simulation of \cite{kw} $10^4$ pencil beam realizations,
recording for each pointing the number count $N$
of galaxies at $6.5\le z\le 7.5$ down to $M_{1500}=-17.5$ and $-19.3$ for the
HUDF and ERS, respectively. The cosmic variance is computed as
$\sigma^2_{CV}=\sigma^2_M-1/<N>$, where the first term is the measured
variance and the second term is the poissonian contribution.
We found that the expected uncertainty due to cosmic variance is
$\sigma_{CV}/<N>=\sim 34\%$ and $\sim 38\%$ for the HUDF and ERS field,
respectively. Combining the two fields, the resulting cosmic variance
yields $\sim 28\%$. These results are in agreement with the predictions
based on \cite{somerville04}, extrapolated to $z\sim 7$.

As shown in \cite{robertson}, since the errors on $M^\ast$, $\Phi^\ast$,
and $\alpha$ are correlated, this does not translate directly in an
additional $\sim 30\%$ uncertainties in the LF parameters.
However, the uncertainties on the luminosity
density or on the UV ionizing photons depends mainly on the total number
of galaxies, so the $\pm 30\%$ additional uncertainties
can be applied to these quantities to
take into account the effects of cosmic variance.

%__________________________________________________________________

\section{Summary}

We have analysed two WFC3 fields, the HUDF and ERS surveys.
More than 40 bona fide z-dropout galaxy candidates have been selected
on these two areas, down to an observed magnitude of $J\sim 29$ (AB).
The depth of ACS images observed with bluer filters (BVI) ensures a
robust and well controlled selection of the high-z galaxy sample,
cleaning the candidates from the large number of interlopers
(i.e. low-z early galaxies, dusty starbursts or cool stars).

We have explored different color criteria for the selection of $z\sim
7$ galaxies, picking up three methods from the wealth of methods
proposed recently in the literature. The numbers of the
resulting candidates are very different, with 13, 6, and 20 candidates
following the A, B, and C criteria (described in Sect. 3) in the HUDF
field, while in the ERS we find 14, 6, and 22 z-dropout candidates,
respectively. We have shown here that, applying these different
methods both to the observed surveys and to the simulated samples, the
resulting LFs are similar.

On the other hand, the output LF depends critically on the
morphological templates used
to simulate the completeness of the observed candidates. In Section 4 we
have proposed 10 different options for a reliable simulation, spanning
from point like objects to very extended galaxies, both observed and
synthetic. The resulting LF is extremely sensitive to the adopted
morphological templates, especially the parameter $\alpha$ governing the
steepness of the faint end. We have shown that the steep LF found by
\cite{bouwens10c} is possibly due to the choice of a set of templates with
typical half light radius larger than the one of the observed
candidates at z=7. If a more compact template is adopted (as used in
\cite{mclure09} and \cite{wilkins2}), a milder slope is derived.

The resulting best fit to the $z\sim 7$ LF is
$Log(\Phi^\ast)=-3.13^{+0.33}_{-0.53}$, 
$M^\ast=-20.14^{+0.40}_{-0.54}$, and $\alpha=-1.65^{+0.71}_{-0.52}$.
This implies a luminosity density (down to an absolute magnitude of -10) of
$\rho_{UV}(-10)=9.2^{+69.8}_{-5.4}\cdot 10^{25} erg/s/Hz/Mpc^3$
and a star formation rate density of
$SFRD(-10)=0.017^{+0.133}_{-0.010} M_\odot/yr/Mpc^3$, if a compact morphology
is used (template 5).
If a more extended morphology is adopted, the resulting best fit to the
$z\sim 7$ LF is
$Log(\Phi^\ast)=-3.22^{+0.36}_{-0.88}$, 
$M^\ast=-20.30^{+0.36}_{-0.82}$, and $\alpha=-2.02^{+0.51}_{-0.59}$.
This implies a luminosity density (down to an absolute magnitude of -10) of
$\rho_{UV}(-10)=4.5^{+436}_{-3.6}\cdot 10^{26} erg/s/Hz/Mpc^3$
and a star formation rate density of
$SFRD(-10)=0.085^{+8.28}_{-0.068} M_\odot/yr/Mpc^3$.

The contribution of z=7 LBGs to the hydrogen ionizing photon density
$\dot{N}_{ion}$ depends on the last two parameters, and the threshold to
keep the Universe ionized at $z\sim 7$ depends also on $f_{esc}$, the escape
fraction of ionizing photons, and on the clumpiness of the IGM, $C_{HII}$.
Limiting the clumpiness
to $C_{HII}\ge 1$, we thus have $f_{esc}\ge 0.17$ in order to have the
Universe reionized by z=7. Inferring a small clumping factor for the IGM
at high-z ($C_{HII}\le 3$ at $z>6$, Bolton \& Haehnelt 2007), an escape
fraction of $f_{esc}\sim 0.5$ is enough for the reionization of the IGM
(but not for the more clumpy galaxy formation regions), assuming a LF
derived from template 5 (compact morphology).
If the IGM was reheated at $z\ge 9$, the observed population of LBGs
at z=7 might be enough to keep the IGM ionized, given that the escape
fraction is larger than 50\%, since a clumpiness of $C_{HII}\sim 2-3$ is
expected, according to recent theoretical predictions (\cite{pawlik09}).

In the near future, two large surveys will provide further information
and better statistics to the $z\sim 7$ related studies: the ground based
wide NIR survey
ULTRAVISTA\footnote{http://www.strw.leidenuniv.nl/$\sim$ultravista/}
(Ultra Deep Survey with the VISTA telescope)
on the COSMOS field and the deep space based survey
CANDELS\footnote{http://candels.ucolick.org/} (Cosmic Assembly Near
Infra-red Deep Extragalactic Legacy Survey).
In particular, the CANDELS survey will have a deep impact on the z=7
study since it will be able to clean out the present high-z sample from
variable objects, to reduce significantly the cosmic variance,
to sample in detail the break of the LF at $M_{1500}\sim -20$, and to provide
a detailed distribution of the half light radii of the candidates, strongly
needed to derive a reliable incompleteness function from the simulations.

In the meanwhile, spectroscopic confirmations of z=7 (and beyond)
candidates with optical/NIR instrumentations from the ground
(\cite{fontana10,vanzella711}), with JWST from space or through sub-mm emission
lines with ALMA will open new frontiers in the study of the first galaxies.

%__________________________________________________________________

\begin{acknowledgements}
Based on observations made with the NASA/ESA Hubble Space Telescope,
obtained from the data archive at the Space Telescope Institute. STScI
is operated by the association of Universities for Research in
Astronomy, Inc. under the NASA contract NAS 5-26555.
Observations were also carried out using the Very Large Telescope at
the ESO Paranal Observatory under Programme IDs LP181.A-0717,
LP168.A-0485, ID 170.A-0788, ID 181.A-0485, ID 283.A-5052 and the ESO Science
Archive under Programme IDs 67.A-0249, 71.A-0584, 73.A-0564, 68.A-0563,
69.A-0539, 70.A-0048, 64.O-0643, 66.A-0572, 68.A-0544, 164.O-0561,
163.N-0210, and 60.A-9120. We acknowledge partial financial support from
ASI.
\end{acknowledgements}


\begin{thebibliography}{}

\bibitem[Becker et al. 2007]{becker07} Becker, G. D., Rauch, M., Sargent,
W. L. W. 2007, ApJ, 662, 72

\bibitem[Beckwith et al. 2006]{udf} Beckwith, S. V. W., Stiavelli, M.,
Koekemoer, A. M., et al. 2006, \aj, 132, 1729

\bibitem[Bertin \& Arnouts 1996]{sex}
Bertin, E. \& Arnouts, S. 1996, \aaps, 117, 393

\bibitem[Bolton \& Haehnelt 2007]{Bolton2007}
Bolton, J.~S. \& Haehnelt, M.~G. 2007, \mnras, 382, 325

\bibitem[Bouwens et al. 2004]{bouwens04}
Bouwens, R. J., Illingworth, G. D., Blakeslee, J. P., Broadhurst, T. J.,
Franx, M., 2004, ApJ, 611, L1

\bibitem[Bouwens \& Illingworth 2006]{bouwens06}
Bouwens, R. J. \& Illingworth, G. D., 2006, Nature, 443, 189

\bibitem[Bouwens et~al. 2007]{bouwens07}
{Bouwens}, R.~J., {Illingworth}, G.~D., {Franx}, M., \& {Ford}, H. 2007, \apj,
670, 928

\bibitem[Bouwens et al. 2008]{bouwens08} Bouwens, R.~J., Illingworth, G.~D.,
Franx, M., \& Ford, H. 2008, \apj, 686, 230

\bibitem[Bouwens et al. 2010a]{bouwens10}
Bouwens, R. J., Illingworth, G. D., Oesch, P. A., et al. 2010a, \apj, 709, L133

\bibitem[Bouwens et al. 2010b]{bouwens10nic}
Bouwens, R. J., Illingworth, G. D., Gonzalez, V., et al. 2010b, \apj, 725, 1587

\bibitem[Bouwens et al. 2011]{bouwens10c}
Bouwens, R. J., Illingworth, G. D., Oesch, P. A., et al. 2011, arXiv:1006.4360

\bibitem[Bridge et al. 2010]{bridge10}
Bridge, C. R., Teplitz, H. I., Siana, B., et al. 2010, ApJ, 720, 465

\bibitem[Bruzual 2007]{Bruzual2007}
{Bruzual}, A.~G. 2007, in IAU Symposium, Vol. 241, IAU Symposium,
ed. A.~{Vazdekis} \& R.~F. {Peletier}, 125--132

\bibitem[Bunker et al. 2010]{bunker09} Bunker, A., Wilkins, S., Ellis, R.,
et al. 2010, MNRAS, 409, 855

\bibitem[Calzetti et~al. 2000]{Calzetti2000}
{Calzetti}, D., {Armus}, L., {Bohlin}, R.~C., {et~al.} 2000, \apj, 533, 682

\bibitem[Capak et al. 2011]{capak} Capak, P., Mobasher, B., Scoville, N. Z.,
et al. 2011, ApJ, 730, 68

\bibitem[Castellano et al. 2010a]{castellano09} Castellano, M., Fontana, A.,
Boutsia, K., et al., 2010, \aap, 511, 20; C10a

\bibitem[Castellano et al. 2010b]{castellano10} Castellano, M., Fontana, A.,
Paris, D., et al. 2010, \aap, 524, 28; C10b

\bibitem[Cen 2003]{cen03} Cen, R., 2003, ApJ, 591, 12

\bibitem[Cen 2010]{cen10} Cen, R., 2010, ApJL submitted, arXiv:1007.0704

\bibitem[Cowie et al. 2010]{cowie10}
Cowie, L. L., Barger, A. J., \& Hu, E. M. 2010, ApJ, 711, 928

\bibitem[Dow-Hygelund et~al. 2007]{Dow2007}
{Dow-Hygelund}, C.~C., {Holden}, B.~P., {Bouwens}, R.~J., {et~al.} 2007, \apj,
660, 47

\bibitem[Dunkley et al. 2009]{dunkley09} Dunkley, J., Komatsu, E., Nolta,
M. R., et al. 2009, ApJS, 180, 306

\bibitem[Fan et al. 2006]{fan06} Fan, X., Strauss, M. A., Becker, R. H.
et al. 2006, AJ, 132, 117

\bibitem[Ferguson et al. 2004]{ferguson04} Ferguson, H. C., Dickinson, M.,
Giavalisco, M., et al. 2004, ApJ, 600, L107

\bibitem[Finkelstein et al. 2010]{finkelstein} Finkelstein, S. L.,
Papovich, C. Giavalisco, M., et al. 2010, ApJ, 719, 1250

\bibitem[Fontana et al. 2010]{fontana10} Fontana, A., Vanzella, E.,
Pentericci, L., et al., 2010, \apj, 725, L205

\bibitem[Giavalisco et~al. 2004]{giavalisco04}
{Giavalisco}, M., {Dickinson}, M., {Ferguson}, H.~C., {et~al.} 2004, \apj,
600, L103

\bibitem[Gnedin \& Fan 2006]{gnedin06} Gnedin, N. Y. \& Fan, X. 2006,
ApJ, 648, 1

\bibitem[Gnedin \& Ostriker 1997]{gnedin} Gnedin, N. Y. \& J. P. Ostriker 1997
\apj, 486, 581

\bibitem[Hickey et al. 2010]{hickey}
{Hickey}, S., {Bunker}, A., {Jarvis}, M.~J., {Chiu}, K., \& {Bonfield}, D.
2010, MNRAS, 404, 212

\bibitem[Kitzbichler \& White 2007]{kw}
{Kitzbichler}, M.~G. \& {White}, S.~D.~M. 2007, \mnras, 376, 2

\bibitem[Koekemoer et al. 2002]{multidrizzle}
Koekemoer, A. M., Fruchter, A. S. Hook, R., Hack, W., 2002,
HST Calibration Worskhop (eds. S. Arribas, A. Koekemoer,
B. Whitmore, STScI: Baltimore), 337

\bibitem[Komatsu et al. 2011]{komatsu10} Komatsu, E., Smith, K. M.,
Dunkley, J., et al. 2011, ApJS, 192, 18

\bibitem[Labb\'e et al. 2006]{labbe06}
Labb\'e, I., Bouwens, R., Illingworth, G. D., Franx, M., 2006, \apj, 649, L67

\bibitem[Madau 1995]{Madau1995} {Madau}, P. 1995, \apj, 441, 18

\bibitem[Madau et al. 1998]{Madau1998} {Madau}, P., Pozzetti, L.,
\& Dickinson, M., 1998, \apj, 498, 106

\bibitem[Madau et~al. 2004]{Madau2004}
{Madau}, P., {Rees}, M.~J., {Volonteri}, M., {Haardt}, F., \& {Oh}, S.~P.
2004, \apj, 604, 484

\bibitem[Mannucci et~al. 2007]{Mannucci2007}
{Mannucci}, F., {Buttery}, H., {Maiolino}, R., {Marconi}, A.,
\& {Pozzetti}, L. 2007, \aap, 461, 423

\bibitem[McLure et al. 2009]{mclure09}
{McLure}, R.~J., {Cirasuolo}, M., {Dunlop}, J.~S., {Foucaud}, S., \& {Almaini},
O., 2009, \mnras, 395, 2196

\bibitem[McLure et al. 2010]{mclure10} McLure, R. J., Dunlop, J. S.,
Cirasuolo, M., et al. 2010, MNRAS, 403, 960

\bibitem[Meiksin 2009]{meiksin09} Meiksin, A. A., 2009, RvMP, 81, 1405

\bibitem[Meurer et al. 1999]{Meurer1999}
{Meurer}, G.~R., {Heckman}, T.~M., \& {Calzetti}, D. 1999, \apj, 521, 64

\bibitem[Oesch et al. 2007]{hudf05} Oesch, P. A., Stiavelli, M.,
Carollo, C. M., et al. 2007, \apj, 671, 1212

\bibitem[Oesch et al. 2009]{hudf09} Oesch, P. A., Carollo, C. M.,
Stiavelli, M., et al. 2009, \apj, 690, 1350

\bibitem[Oesch et al. 2010]{oesch09} Oesch, P. A., Bouwens, R. J.,
Illingworth, G. D., et al. 2010, \apj, 709L, 16

\bibitem[Oesch et al. 2010b]{oesch09b} Oesch, P. A., Bouwens, R. J.,
Carollo, C. M., et al. 2010, \apj, 709, L21

\bibitem[Ouchi et al. 2009]{ouchi} Ouchi, M., Mobasher, B., Shimasaku, K.,
et~al. 2009, \apj, 706, 1136

\bibitem[Overzier et~al. 2009]{Overzier2009}
{Overzier}, R.~A., {Guo}, Q., {Kauffmann}, G., {et~al.} 2009, \mnras, 394, 577

\bibitem[Pawlik et al. 2010]{pawlik09}
Pawlik, A. H., Schaye, J., van Scherpenzeel, E. 2010, ASPC, 432, 230

\bibitem[Robertson 2010]{robertson} Robertson, B. E., 2010, \apj, 713, 1266

\bibitem[{{Schechter}(1976)}]{Schechter1976}
{Schechter}, P. 1976, \apj, 203, 297

\bibitem[Siana et al 2010]{siana10}
Siana, B., Teplitz, H. I., Ferguson, H. C., et al. 2010, ApJ, 723, 241

\bibitem[Somerville et al. 2004]{somerville04}
Somerville, R. S., Lee, K., Ferguson, H. C., et al. 2004, ApJ, 600, L171

\bibitem[Songaila \& Cowie 2010]{songaila10}
Songaila, A. \& Cowie, L. L. 2010, ApJ, 721, 1448

\bibitem[Stanway et~al. 2008]{Stanway2008}
{Stanway}, E.~R., {Bremer}, M.~N., {Squitieri}, V., {Douglas}, L.~S., \&
{Lehnert}, M.~D. 2008, \mnras, 386, 370

\bibitem[Vanzella et al. 2009]{vanzella}
Vanzella, E., Giavalisco, M., Dickinson, M., et al. 2009, ApJ, 695, 1163

\bibitem[Vanzella et al. 2010]{vanzella10}
Vanzella, E., Giavalisco, M., Inoue, A., et al. 2010, ApJ, 725, 1011

\bibitem[Vanzella et al. 2011]{vanzella711}
Vanzella, E., Pentericci, L., Fontana, A., et al. 2011, ApJ, 730, L35

\bibitem[Venkatesan et~al. 2003]{Venkatesan2003}
{Venkatesan}, A., {Tumlinson}, J., \& {Shull}, J.~M. 2003, \apj, 584, 621

\bibitem[Wilkins et~al. 2010]{wilkins}
{Wilkins}, S.~M., {Bunker}, A.~J., {Ellis}, R.~S., {et~al.}
2010, MNRAS, 403, 938

\bibitem[Wilkins et~al. 2011]{wilkins2}
{Wilkins}, S.~M., {Bunker}, A.~J., {Lorenzoni}, S., {et~al.}
2011, MNRAS, 411, 23

\bibitem[Windhorst et al 2002]{windhorst02} Windhorst, R. A., Cohen, S.,
Jansen, R., et al. 2002, AAS, 201, 3207

\bibitem[Windhorst et al 2011]{windhorst} Windhorst, R. A., Cohen, S. H.,
Hathi, N. P., et al. 2011, ApJS, 193, 27

\bibitem[Yan et~al. 2010]{yan}
{Yan}, H., {Windhorst}, R., {Hathi}, N., {et~al.} 2010, RAA, 10, 867

\bibitem[Yan et~al. 2011]{yan11}
Yan, H., Yan, L., Zamojski, M. A., et al. 2011, \apj, 728, 22

\end{thebibliography}
\end{document}